\documentclass[10pt,aps,prd,twocolumn,showpacs,superscriptaddress,eqsecnum,longbibliography,nofootinbib]{revtex4-2}
\usepackage{mathrsfs,amsmath,amsthm,latexsym,amssymb,amsfonts,epsfig,cancel,enumerate,graphicx,subcaption,txfonts}
\usepackage{multirow}

\usepackage{xcolor}
\usepackage[colorlinks,linkcolor=blue,citecolor=blue,urlcolor=blue,plainpages=false,pdfstartview=FitH]{hyperref}
\usepackage{appendix}
\allowdisplaybreaks
\newcommand{\GZU}{School of Physics, Guizhou University, Guiyang 550025, China}

\begin{document}

\title{Motion of spinning particles around a quantum-corrected black hole without Cauchy horizons}

\author{Jiawei Chen}
\email{gs.chenjw23@gzu.edu.cn}
\affiliation{\GZU}

\author{Jinsong Yang}
\thanks{Corresponding author}
\email{jsyang@gzu.edu.cn}
\affiliation{\GZU}

\begin{abstract}
In this paper, we investigate the motion of spinning particles around a covariant quantum-corrected black hole without a Cauchy horizon within the framework of effective quantum gravity, and examine the influence of quantum gravitational effects on the motion of these spinning particles. First, we employ the Mathisson-Papapetrou-Dixon equations to derive the 4-momentum and 4-velocity of spinning particles, and introduce the effective potential for radial motion using the components of the 4-momentum. We find that an increase in the quantum parameter $\zeta$ leads to a decrease in the effective potential, while the spin $S$ significantly affects the magnitude of the effective potential. Then, through the effective potential, we investigate the properties of circular orbits and the innermost stable circular orbit, and discuss the timelike condition that spinning particles must satisfy when moving around the black hole. Finally, we study the trajectories of spinning particles on bound orbits around the quantum-corrected black hole and compare them with those around other covariant quantum-corrected black holes. The results show that the trajectories of spinning particles in this quantum-corrected black hole model are weakly influenced by $\zeta$, making them almost indistinguishable from those in the Schwarzschild black hole, but they can be distinguished from other covariant quantum-corrected models under certain initial conditions. These results contribute to our understanding of black hole properties under quantum corrections.
\end{abstract}

\maketitle

\section{Introduction}

The spacetime singularities~\cite{Penrose:1964wq,Hawking:1970zqf} in general relativity (GR) and its incompatibility with quantum physics~\cite{Surya:2019ndm} have motivated the development of modified gravity theories and quantum gravity approaches to address these fundamental issues. Among these, loop quantum gravity (LQG), as a non-perturbative and background-independent theory of quantum gravity, has attracted widespread attention. In addressing spacetime problems, LQG discretizes the continuous spacetime of GR into fundamental units~\cite{Rovelli:2011eq}, thereby avoiding curvature divergence. Although significant research progress has been achieved in LQG~\cite{Rovelli:1997yv,Ashtekar:2005qt,Ashtekar:2004eh,Ashtekar:2013hs,Han:2005km,Modesto:2008im,Perez:2017cmj,Ashtekar:2018lag,Bodendorfer:2019cyv,Kelly:2020uwj,Gan:2020dkb,Sartini:2020ycs,Song:2020arr,Zhang:2020qxw,Zhang:2021wex,Lewandowski:2022zce}, the theory still faces several challenges, such as the issue of covariance~\cite{Bojowald:2015zha}. The principle of general covariance requires the mathematical expressions of physical laws to remain invariant under any coordinate transformation. However, it remains unclear whether covariance is preserved in the effective Hamiltonian theory derived from canonical quantum gravity. This issue has sparked extensive discussion in the community~\cite{Bojowald:2015sta,BenAchour:2017jof,Bojowald:2020unm,Gambini:2022dec,Han:2022rsx,Li:2023axl}. Recently, new progress has been made in the study of covariance. Within the framework of LQG-inspired effective quantum gravity, three static quantum-corrected black hole (BH) solutions satisfying general covariance have been obtained~\cite{Zhang:2024khj,Zhang:2024ney}, following the rigorous derivation of the conditions for general covariance in static spherically symmetric gravity. Subsequently, this framework was extended to the electromagnetic vacuum case with a cosmological constant, successfully obtaining a family of charged quantum-corrected BH models with a cosmological constant that preserve covariance~\cite{Yang:2025ufs}. This has stimulated extensive research on these covariant quantum-corrected BH models~\cite{Konoplya:2024lch,Liu:2024soc,Liu:2024wal,Zhu:2024wic,Wang:2024iwt,Ban:2024qsa,Lin:2024beb,Shu:2024tut,Liu:2024iec,Bojowald:2024ium,Konoplya:2025hgp,Chen:2025ifv,Lutfuoglu:2025hwh,Chen:2025aqh,Al-Badawi:2025rcq,Zhang:2025ccx,Sekhmani:2025bsi}.

The successful detection of gravitational waves (GWs)~\cite{LIGO:2017dbh} and the release of images of supermassive BHs~\cite{EventHorizonTelescope:2019dse} have not only provided strong evidence for GR but also established a critical testing ground for various alternative theories of gravity. Through systematic studies of photon motion around BHs, one can precisely calculate the deflection angles of light rays and extract characteristic features of BH from shadows, thereby offering potential means to distinguish between quantum-corrected BH models and classical BH spacetime structures. On the other hand, investigating the properties of BHs by studying the motion of timelike particles along geodesics also serves as an effective approach, aiding in understanding the trajectories of matter around BHs and the emission characteristics of GWs~\cite{Hughes:2000ssa}. However, under conditions of strong fields or high angular momentum, the geodesic approximation for test particles may no longer be valid, especially when particles possess intrinsic spin. In such cases, the spin-curvature coupling effect between the particle and the gravitational field significantly influences its motion~\cite{Jefremov:2015gza}. Therefore, extending the study from spinless particles to spinning particles and systematically investigating their dynamical properties may become a crucial step in further exploring the physical mechanisms of strong gravitational fields.

Particles with intrinsic spin no longer follow geodesic motion in a gravitational field; instead, their trajectories are governed by the Mathisson-Papapetrou-Dixon (MPD) equations~\cite{Mathisson:1937zz,Papapetrou:1951pa,Dixon:1970zza,Dixon:1970zz,Dixon:1974xoz}. In the pole-dipole approximation, these equations fully incorporate the coupling effect between spin and spacetime curvature. In recent years, extensive research has been conducted on the dynamical behavior of spinning particles in various BH spacetimes~\cite{Jefremov:2015gza,Zhang:2017nhl,Toshmatov:2019bda,Zhang:2022qzw,Rakhimova:2024hzt,Tan:2024hzw,Skoupy:2024uan}, including quantum-corrected BH spacetimes~\cite{Alimova:2025izs,Du:2024ujg,Umarov:2025wzm}. Recently, preliminary analyses have been carried out on the motion characteristics of spinning particles around two covariant quantum BHs within the framework of effective quantum gravity~\cite{Du:2024ujg,Umarov:2025wzm}. However, the motion of spinning particles around the third covariant quantum BH model without a Cauchy horizon remains unexplored. This gap motivates us to further investigate the dynamics of spinning particles in this model in the present paper, aiming to reveal the potential influence of its quantum corrections on the motion of spinning particles.

In this study, we investigate the motion of spinning particles around a quantum-corrected BH without a Cauchy horizon and examine the influence of quantum corrections on the motion of these spinning particles. Specifically, we first employ the MPD equations under the pole-dipole approximation, along with supplementary conditions, to derive the equations of motion for spinning particles in this spacetime. Based on the obtained 4-momentum of the spinning particles, we define the effective potential for their radial motion and investigate the effects of spin $S$ and quantum parameter $\zeta$ on this effective potential. Subsequently, using the effective potential, we discuss the circular orbits of spinning particles around this quantum-corrected BH and analyze the dependence of the circular orbit radius on $S$ and $\zeta$. Furthermore, we study the influence of $S$ and $\zeta$ on the innermost stable circular orbit (ISCO) and determine the constraints on $S$ and $\zeta$ for spinning particles at the ISCO to satisfy the timelike condition. Finally, we present the trajectories of spinning particles around this covariant quantum-corrected BH and provide a brief comparison with the trajectories of spinning particles around two other covariant quantum-corrected BHs.

The structure of this paper is organized as follows. In Sec.~\ref{section2}, we briefly recall the quantum-corrected BH without a Cauchy horizon. Then, under the pole-dipole approximation, we use the MPD equations to study the motion of spinning particles and define the effective potential for their radial motion. In Sec.~\ref{section3}, based on the effective potential, we investigate the circular orbits and ISCO of spinning particles around the quantum-corrected BH, and determine the parameter constraints for spinning particles at the ISCO to satisfy the timelike condition. Sec.~\ref{seciton4} presents the corresponding trajectories of spinning particles around the quantum-corrected BH and compares them with those around other covariant quantum-corrected BHs. Finally, a summary is provided in Sec.~\ref{section5}. Throughout this paper, we adopt geometric units with $G=c=1$ and set the BH mass $M=1$ for calculations.

\section{Equations of Spinning Particles in Quantum-Corrected Spacetime}\label{section2}

\subsection{A brief review of quantum-corrected BHs without Cauchy horizons}

Recently, within the framework of effective quantum gravity, the issue of general covariance has been successfully resolved, and the covariance condition equations have been proposed~\cite{Zhang:2024khj}. Subsequently, by solving these covariance equations, a covariant quantum-corrected BH solution without a Cauchy horizon, which contains quantum parameters, was further derived~\cite{Zhang:2024ney}. In this section, we provide a brief review of this quantum-corrected BH. In Schwarzschild coordinates, its line element can be expressed in the general form as:
\begin{equation}
	{\rm d}s^2=g_{tt}{\rm d}t^2+g_{rr}{\rm d}r^2+g_{\theta \theta} {\rm d}\theta^2+g_{\phi \phi} {\rm d}\phi^2,\label{metric}
\end{equation}
where
\begin{equation}
	\begin{aligned}
		g_{tt}&=- \left[1-(-1)^n\frac{r^2}{\zeta^2}\arcsin\left(\frac{2M\zeta^2}{r^3}\right)-\frac{n\pi r^2}{\zeta^2} \right],\\
		g_{rr}&= \frac{1}{-g_{tt} (1-\frac{4M^2\zeta^4}{r^6})},\\
		g_{\theta \theta}&=r^2, \\
		g_{\phi \phi}&=r^2 \sin^2{\theta}.
	\end{aligned}
\end{equation}
Here, $M$ represents the BH mass, $\zeta$ denotes the quantum parameter, and $n$ is an arbitrary integer. Moreover, if and only if $n=0$ and $\zeta \rightarrow 0$, the metric reduces to the Schwarzschild case. Throughout this paper, we adopt the case of $n=0$, and for convenience, we uniformly denote $\zeta=0$ as representing the Schwarzschild BH.

Under the joint constraints of the $\arcsin$ function and the metric component $g_{rr}$ on the BH horizon, we derive the condition that the quantum parameter $\zeta$ must satisfy~\cite{Zhang:2024ney}
\begin{equation}
	\zeta/ M < 2\left( \frac{\pi}{2} \right)^{3/2} \cong 3.94.\label{maxzeta}
\end{equation}
Subsequent parameter selections will be strictly confined to this range.

\subsection{Equations of motion for spinning particles in quantum-corrected spacetime}

We consider a spinning test particle with mass smaller than that of the BH, whose motion in this quantum-corrected BH spacetime is governed by the MPD equations under the pole-dipole approximation~\cite{Mathisson:1937zz,Papapetrou:1951pa,Dixon:1970zza,Dixon:1970zz,Dixon:1974xoz,Jefremov:2015gza,Zhang:2017nhl,Toshmatov:2019bda}
\begin{eqnarray}
	\frac{DP^{a}}{D\tau} &=& -\frac{1}{2} R^{a}_{\ bcd} u^{b} S^{cd}, \label{MPD1}\\
	\frac{DS^{ab}}{D\tau} &=& -P^{b} u^{a} + P^{a} u^{b}\label{MPD2},
\end{eqnarray}
where $P^a$, $u^{a}$, and $\tau$ represent the spinning particle's 4-momentum, 4-velocity along the trajectory, and affine parameter of the trajectory, respectively. Furthermore, $R^{a}_{\ bcd}$ denotes the Riemann tensor, and $S^{ab}$ is the antisymmetric spin tensor. Due to the fact that spinning particles possess more degrees of freedom than the number of equations in Eqs.~\eqref{MPD1} and \eqref{MPD2}, the system cannot be solved directly and requires additional supplementary conditions to be imposed. Since the remaining undetermined degrees of freedom are related to the center of mass of the spinning particle and are observer-dependent, this additional supplementary condition is not unique~\cite{Tan:2024hzw,Umarov:2025wzm}. Here, we adopt the Tulczyjew Spin Supplementary Condition (SSC)~\cite{Jefremov:2015gza,Zhang:2017nhl,Toshmatov:2019bda,Zhang:2022qzw,Rakhimova:2024hzt,Tan:2024hzw,Skoupy:2024uan,Alimova:2025izs,Du:2024ujg,Umarov:2025wzm}
\begin{equation}
	S^{ab}P_{b}=0. \label{condition}
\end{equation}
Then, we can define two conserved quantities using the spin tensor and the 4-momentum
\begin{eqnarray}
	P^{a} P_{a} &=& -m^2, \label{m} \\
	S^{ab} S_{ab} &=& 2 \bar{S}^2. \label{Sbar}
\end{eqnarray}
Here, $m$ and $\bar{S}$ denote the mass and spin of the test particle.

For convenience, we restrict the motion of the spinning particle to the equatorial plane ($\theta = \pi/2$). Hence, $P^{\theta}$ and all $S^{\mu \theta}$ components vanish. The antisymmetric spin tensor has only three independent components: $S^{tr}$, $S^{t\phi}$, and $S^{r\phi}$. Then, from Eq.~\eqref{condition}, we obtain
\begin{eqnarray}
	S^{tr} P_r + S^{t\phi} P_{\phi}&=& 0, \label{S1} \\
	-S^{tr} P_t + S^{r\phi} P_{\phi}&=& 0, \label{S2}\\
	-S^{t\phi} P_t - S^{r\phi} P_{r}&=& 0. \label{S3}
\end{eqnarray}
Combining Eqs.~\eqref{m} and \eqref{Sbar}, we solve Eqs.~\eqref{S1}--\eqref{S3} to obtain
\begin{eqnarray}
	S^{tr}&=&-S^{rt}=S\frac{P_\phi}{F},\\
	S^{t\phi}&=&-S^{\phi t}=-S\frac{P_r}{F},\\
	S^{r\phi}&=&-S^{\phi r}=S\frac{P_{t}}{F},
\end{eqnarray}
where $S \equiv \bar{S}/m$ and the function $F$ can be expressed as
\begin{equation}
	F=\sqrt{-g_{tt} g_{rr} g_{\phi \phi}}. \label{function}
\end{equation}

Furthermore, for a spinning particle, the conserved quantities $\mathcal{C}$ generated by Killing vector fields $\mathcal{\xi}_a$ can be expressed as follows~\cite{Jefremov:2015gza,Zhang:2017nhl,Toshmatov:2019bda,Zhang:2022qzw,Rakhimova:2024hzt,Tan:2024hzw,Skoupy:2024uan}
\begin{equation}
	\mathcal{C}= P^a \mathcal{\xi}_a- \frac{1}{2} S^{ab}\nabla_b \mathcal{\xi}_{a}.
	\label{Eq:Conserved}
\end{equation}
The spacetime described by Eq.~\eqref{metric} possesses two Killing vector fields, $\kappa^a \equiv (\partial /\partial t)^a$ and $\eta^a \equiv (\partial /\partial \phi)^a$, which yield two additional conserved quantities: energy $\bar{E}$ and angular momentum $\bar{J}$
\begin{eqnarray}
	\bar{E}=-\mathcal{C}_t&=&-P^a \kappa_a +\frac{1}{2} S^{ab}\nabla_b \mathcal{\kappa}_{a}\nonumber\\
	&=&-P_t+\frac{1}{2}\frac{S}{F}P_\phi\partial_r g_{tt},\label{E}\\
	\bar{J}=\mathcal{C}_{\phi}&=&P^a \eta_a -\frac{1}{2} S^{ab}\nabla_b \mathcal{\eta}_{a}\nonumber\\
	&=&P_\phi+\frac{1}{2}\frac{S}{F}P_t\partial_r g_{\phi\phi}.\label{J}
\end{eqnarray}
Then we derive the 4-monmentum with the conserved quantities
\begin{eqnarray}
	P_t &=& -\frac{2\,m\,F \left(2\,E\,F - J\,S\,\partial_r g_{tt}\right)} {4 F^2 + S^2 \partial_r g_{\phi\phi}\partial_r g_{tt}}, \label{Pt}\\
	P_\phi &=& \frac{2\,m\,F \left(2\,J\,F + E\,S\,\partial_r g_{\phi \phi} \right)}{4F^2+ S^2 \partial_r g_{\phi\phi}\partial_r g_{tt}},\label{Pphi} \\
	(P^r)^2 &=&-\frac{m^2+g^{tt}P_t^2+g^{\phi\phi}P_\phi^2}{g_{rr}}. \label{pr}
\end{eqnarray}
Note that here $E \equiv \bar{E}/m$ and $J \equiv \bar{J}/m$. We further introduce the orbital angular momentum $L$, defined as $L = J - S$, to distinguish it from the total angular momentum $J$.

With these definitions in place, to further simplify the calculation, we set the affine parameter $\tau$ equal to the coordinate time $t$, so that $u^t = 1$. By substituting the components of the spin tensor $S^{ab}$ into the MPD equations~\eqref{MPD1} and \eqref{MPD2}, we derive the governing equations for $u^r$ and $u^\phi$
\begin{eqnarray}
	\frac{DS^{tr}}{D\tau} &=& P^t u^{r}-P^r=-\frac{S}{2F}g_{\phi a}R^a_{bcd}u^b S^{cd}, \label{v_tr}\\
	\frac{DS^{t\phi}}{D\tau}
	&=& P^t u^{\phi}-P^\phi
	= \frac{S}{2F}g_{r a}R^a_{bcd}u^b S^{cd}. \label{v_tphi}
\end{eqnarray}
Thus we obtain
\begin{eqnarray}
	u^r&=&\frac{P^r-\frac{S}{2F}R_{\phi t\mu\nu}S^{\mu\nu}}{P^t+\frac{S}{2F}R_{\phi r\mu\nu}S^{\mu\nu}},\label{vr}\\
	u^{\phi}&=&\frac{P^\phi+\frac{S}{2F}R_{rt\mu\nu}S^{\mu\nu}}{P^t-\frac{S}{2F}R_{r \phi\mu\nu}S^{\mu\nu}}.\label{vphi}
\end{eqnarray}

\subsection{Effective potential of the radial motion}

When studying the motion of particles in the equatorial plane around a BH, an effective potential is usually introduced to analyze the behavior of their radial motion. Since the radial velocity $u^r$ is proportional to the radial momentum $P^r$, here we define the effective potential for spinning particles in terms of the radial momentum~\cite{Tan:2024hzw,Zhang:2022qzw}
\begin{eqnarray}
	(P^r)^2&=&\left(X E^2+Y E+Z\right)\nonumber\\
	&=& \left(E-\frac{-Y+\sqrt{Y^2-4XZ}}{2X}\right)\nonumber\\
	&\times& \left(E-\frac{-Y-\sqrt{Y^2-4XZ}}{2X}\right),
	\label{Veff}
\end{eqnarray}
with
\begin{align*}
	X &= -\frac{4m^2 F^2 \left(4g^{tt} F^2 + g^{\phi\phi} S^2 (\partial_r g_{\phi\phi})^2\right)}{g_{rr} W}, \\
	Y &= -\frac{16m^2 F^3 J S \left(g^{\phi\phi} \partial_r g_{\phi\phi} - g^{tt} \partial_r g_{tt}\right)}{g_{rr} W}, \\
	Z &= \frac{-m^2}{g_{rr} W} \left[W - 4F^2 \left(g^{tt} J^2 S^2 (\partial_r g_{tt})^2 + 4g^{\phi\phi} J^2 F^2\right)\right], \\
	W &= \left(4F^2 + S^2 \partial_r g_{\phi\phi} \partial_r g_{tt}\right)^2.
\end{align*}
Since the motion of spinning particles should be future-directed, we define the positive root as the effective potential~\cite{Zhang:2022qzw,Tan:2024hzw}
\begin{equation}
	V_{\rm{eff}} = \frac{-Y+\sqrt{Y^2-4XZ}}{2X}.
	\label{effective}
\end{equation}

\begin{figure*}[htb]
	\centering
	\begin{subfigure}{0.33\textwidth}
		\includegraphics[width=6cm, keepaspectratio]{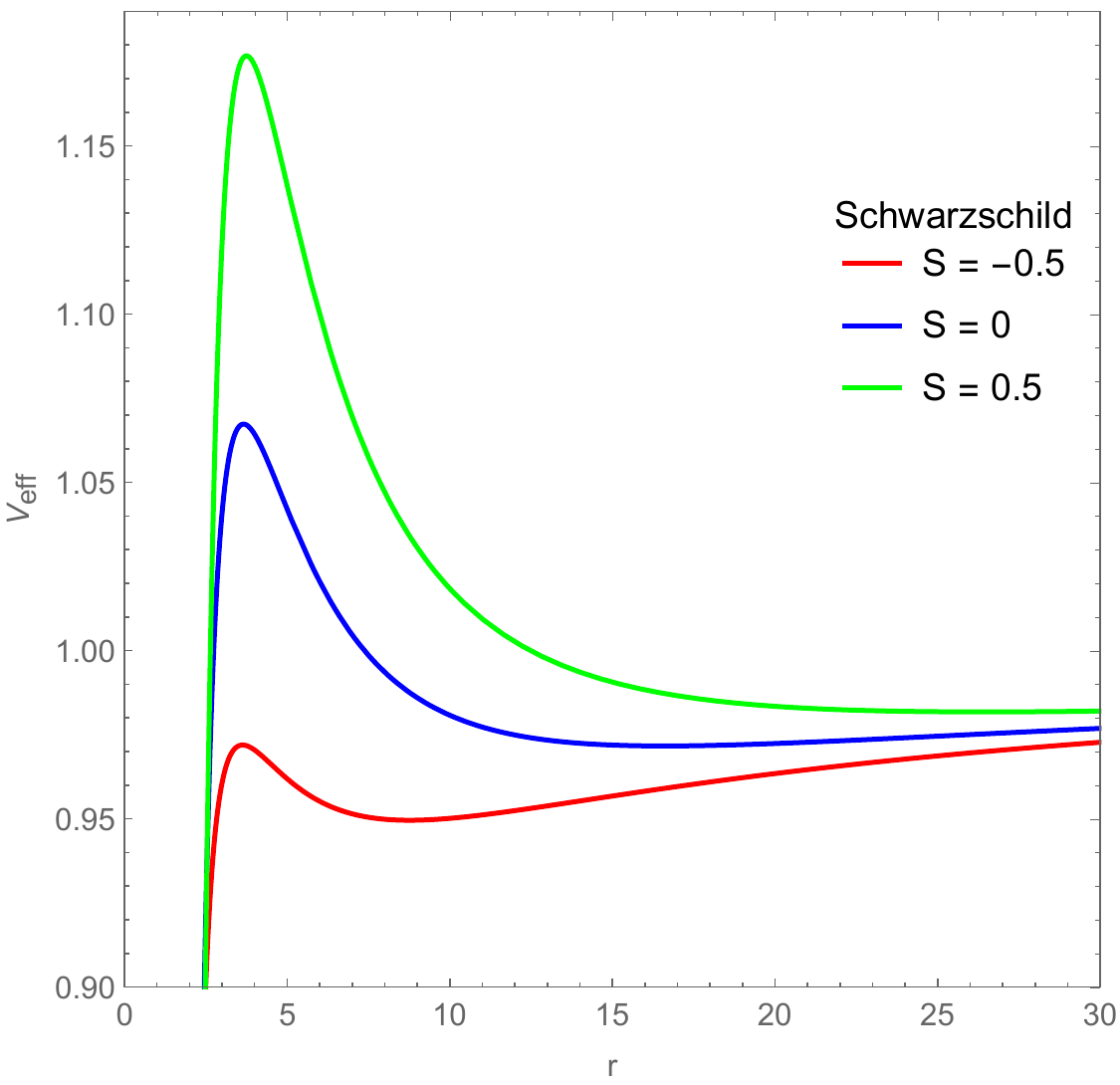}
	\end{subfigure}
	\hfill
	\begin{subfigure}{0.33\textwidth}
		\includegraphics[width=6cm, keepaspectratio]{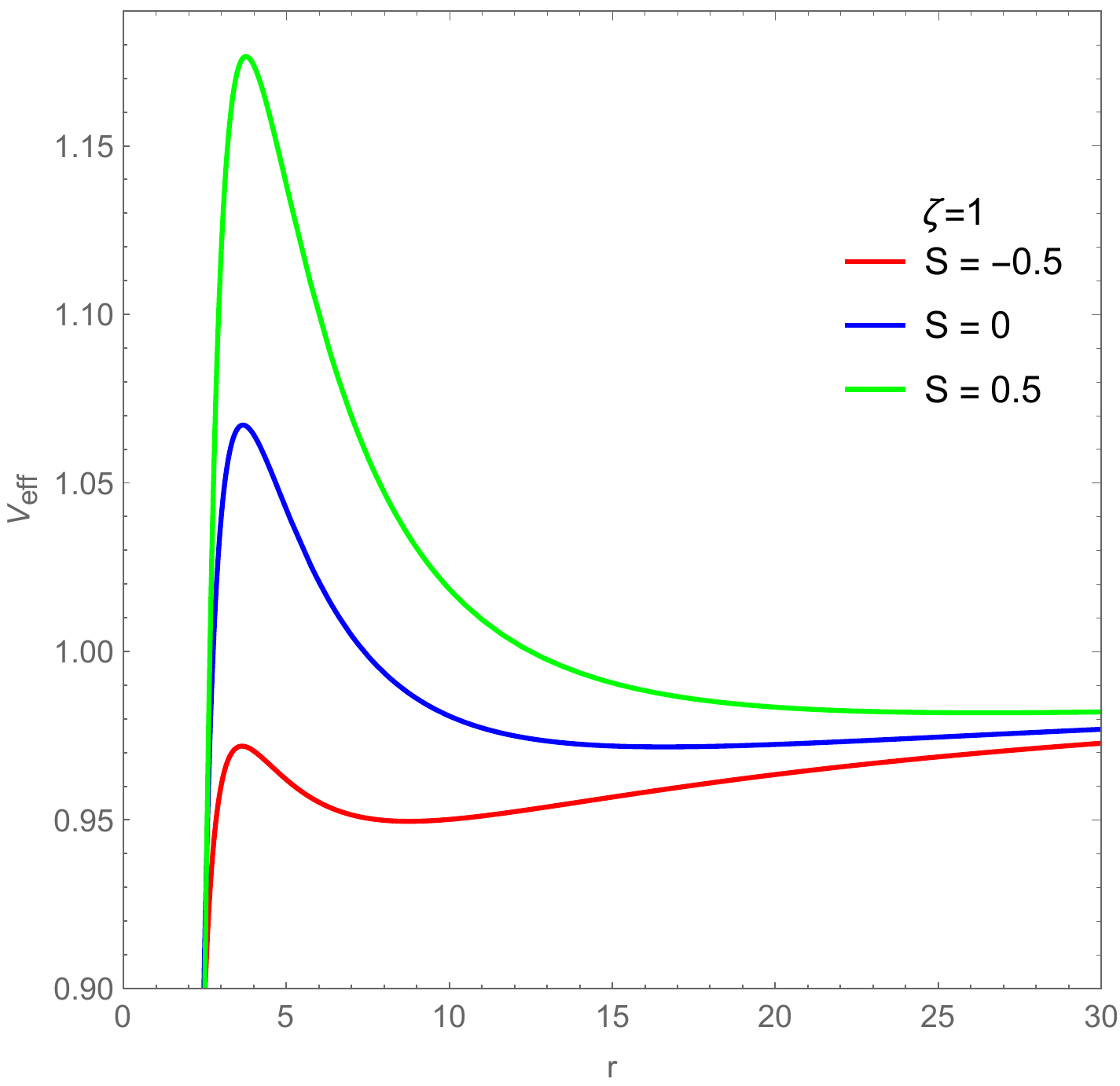}
	\end{subfigure}
	\hfill
	\begin{subfigure}{0.33\textwidth}
		\includegraphics[width=6cm, keepaspectratio]{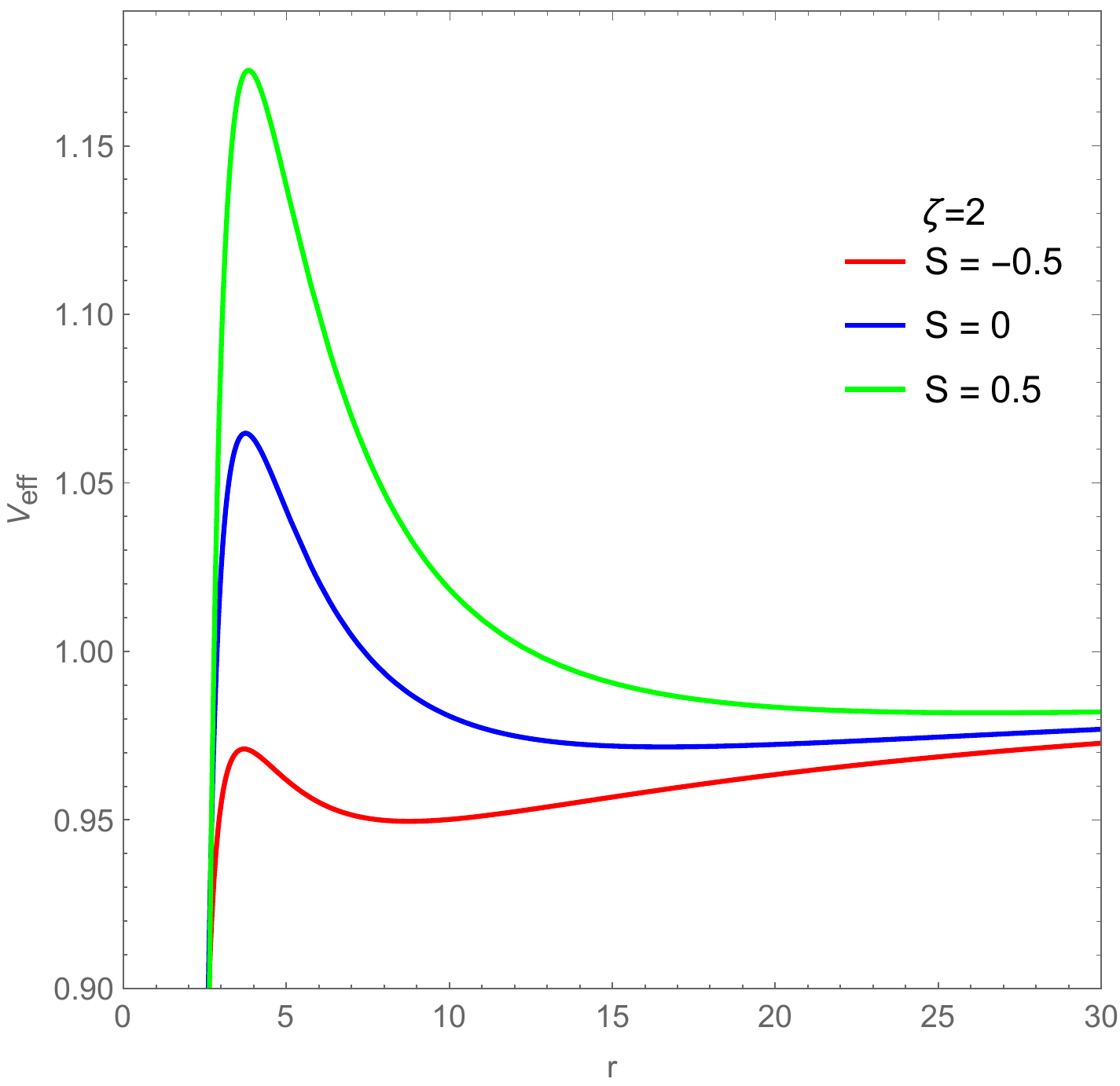}
	\end{subfigure}
	\hfill
	\begin{subfigure}{0.33\textwidth}
		\includegraphics[width=6cm, keepaspectratio]{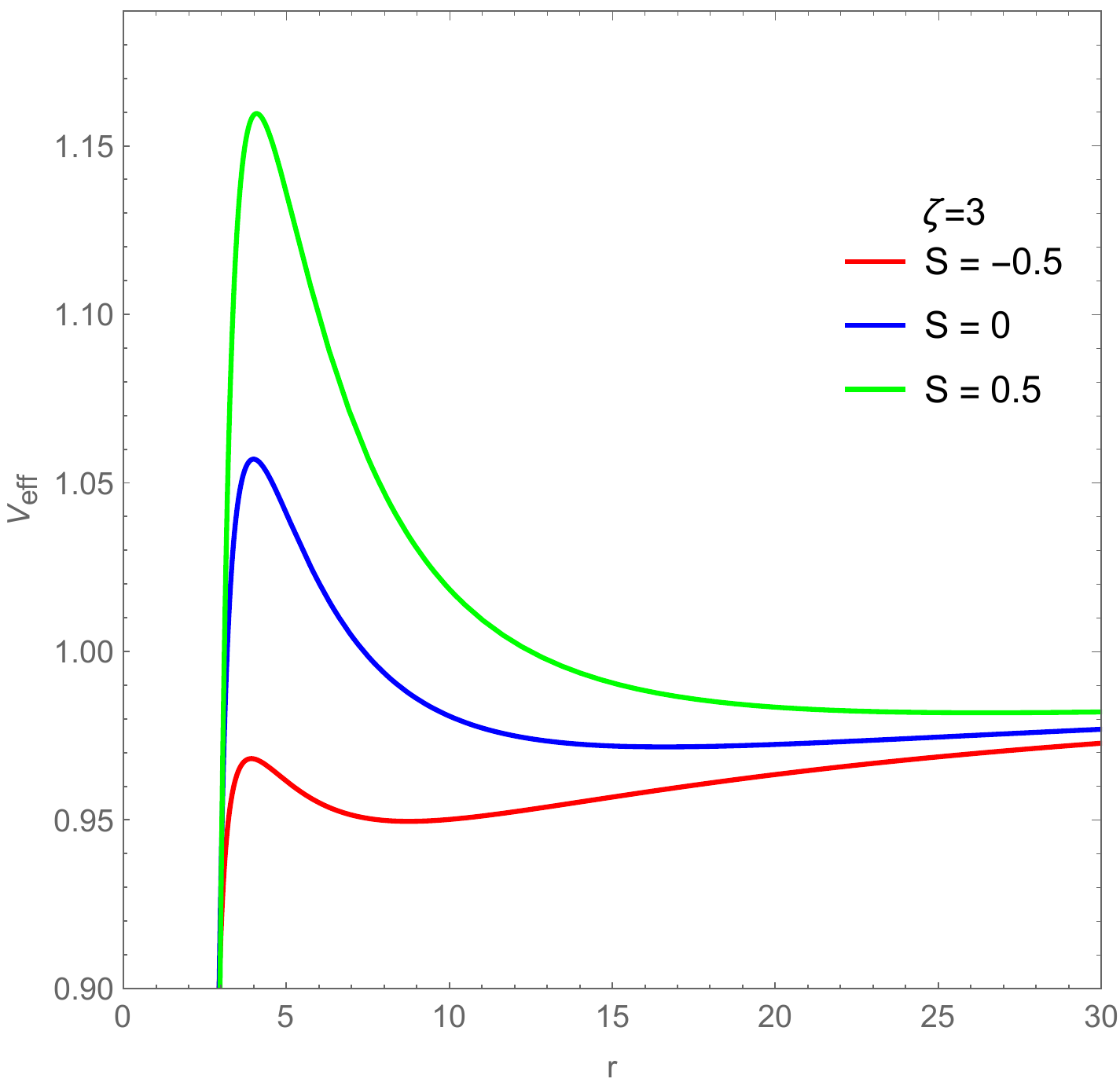}
	\end{subfigure}
	\hfill
	\begin{subfigure}{0.33\textwidth}
		\includegraphics[width=6cm, keepaspectratio]{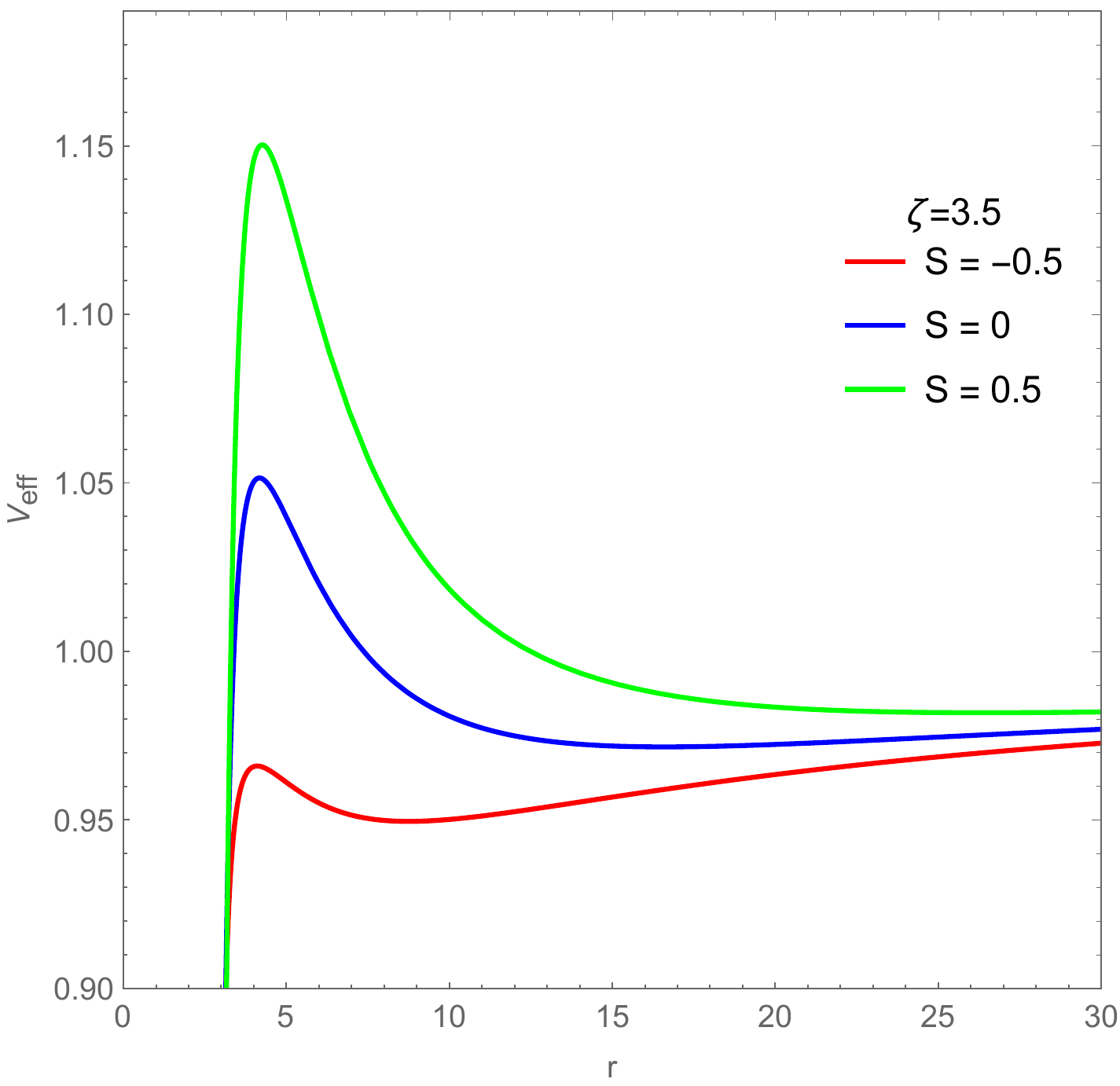}
	\end{subfigure}
	\hfill
	\begin{subfigure}{0.33\textwidth}
		\includegraphics[width=6cm, keepaspectratio]{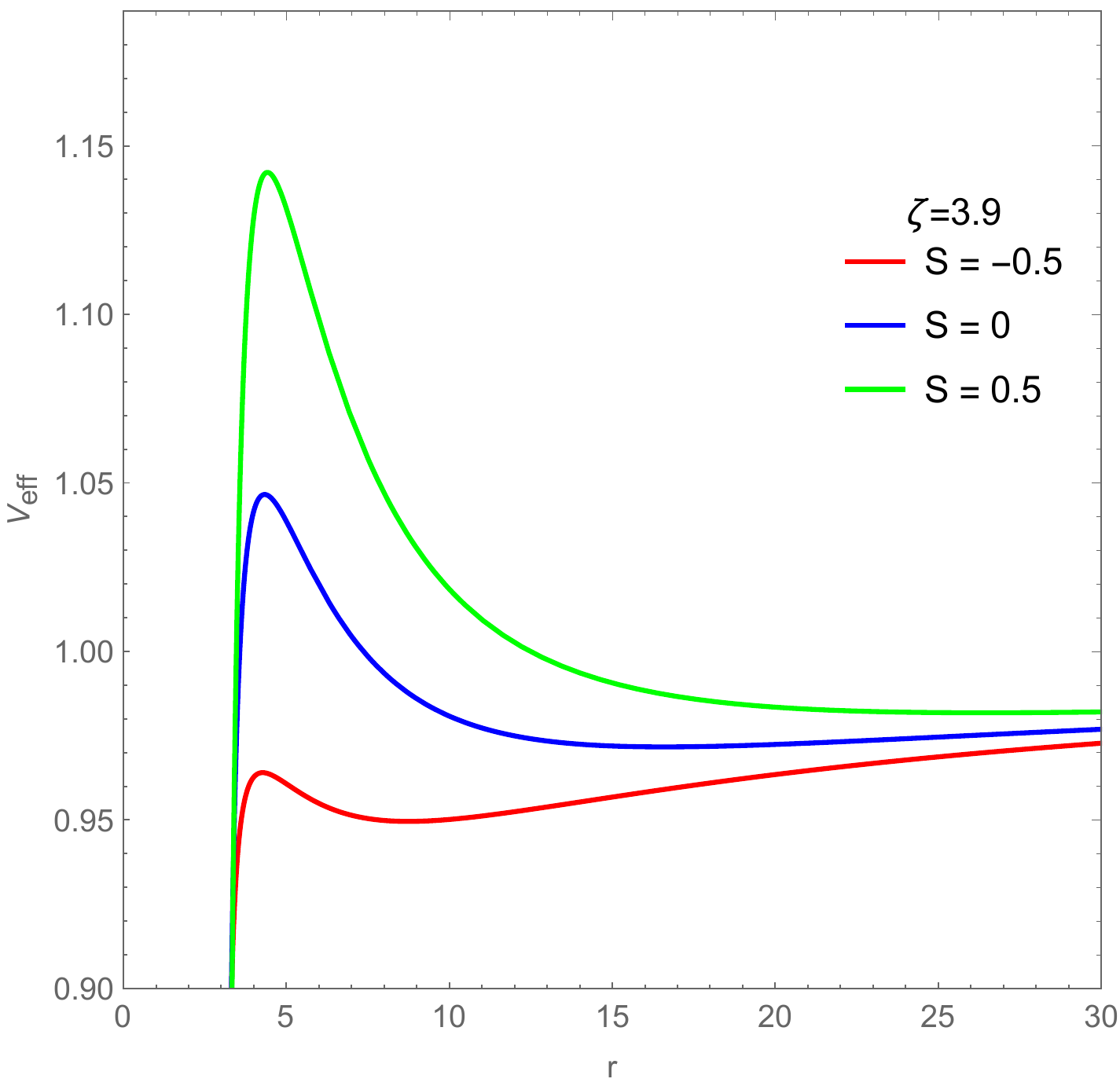}
	\end{subfigure}
	\hfill
	\caption{Variation of the effective potential $V_{\rm eff}$ for different values of the quantum parameter $\zeta$ and spin parameter $S$ at fixed $L=4.5$.}
	\label{fig:veff}
\end{figure*}

In Fig.~\ref{fig:veff}, we show the variation of the effective potential $V_{\rm eff}$ with the radial coordinate $r$ for different values of the quantum parameter $\zeta$ and spin parameter $S$ at a fixed $L=4.5$. It can be seen that for a fixed $\zeta$, an increase in spin $S$ leads to a rise in the effective potential. However, for a fixed $S$, the value of the effective potential gradually decreases as $\zeta$ increases. Moreover, the influence of $S$ on the effective potential is more significant than that of $\zeta$.

Building upon previous studies of the effective potential for spinning particles~\cite{Umarov:2025wzm} around the other two covariant quantum-corrected BHs, we find that the behavior of the effective potential with respect to $\zeta$ (at fixed $S$) differs among the three models. In the first model (denoted as BH-I), the effective potential increases with growing $\zeta$ (regardless of whether $S$ is positive or negative). The second model, referred to as BH-II, has a dependence on $\zeta$ that additionally depends on the sign of $S$, and when $S=0$, the value of $\zeta$ does not affect the effective potential. In contrast, as discussed earlier, the covariant quantum BH model considered here (denoted as BH-III) exhibits a decrease in the effective potential with increasing $\zeta$ (again, regardless of the sign of $S$).

\section{Circular Orbits Around Quantum-Corrected BHs} \label{section3}

\subsection{Circular orbits}

In previous discussions, we have derived the effective potential describing the radial motion of spinning particles. Among the orbits of spinning particles around quantum-corrected BHs, there exists a special type known as circular orbits, which can be defined using the effective potential. For particles in circular orbits, both the radial velocity and radial acceleration vanish, implying that the effective potential must satisfy the conditions:
\begin{equation}
	\begin{split}
		V_\text{eff}= E, \quad
		\frac{{\rm d} V_\text{eff}}{{\rm d}r} = 0.\label{circle}
	\end{split}
\end{equation}
Furthermore, the stability of circular orbits can be determined by $\frac{{\rm d}^2 V_{\rm eff}}{{\rm d} r^2}$. For stable circular orbits, $\frac{{\rm d}^2 V_{\rm eff}}{{\rm d} r^2} \ge 0$, while for unstable orbits, $\frac{{\rm d}^2 V_{\rm eff}}{{\rm d} r^2} < 0$.

\begin{figure*}[htbp]
	\centering
	\begin{subfigure}{0.45\textwidth}
		\includegraphics[width=3in, height=5in, keepaspectratio]{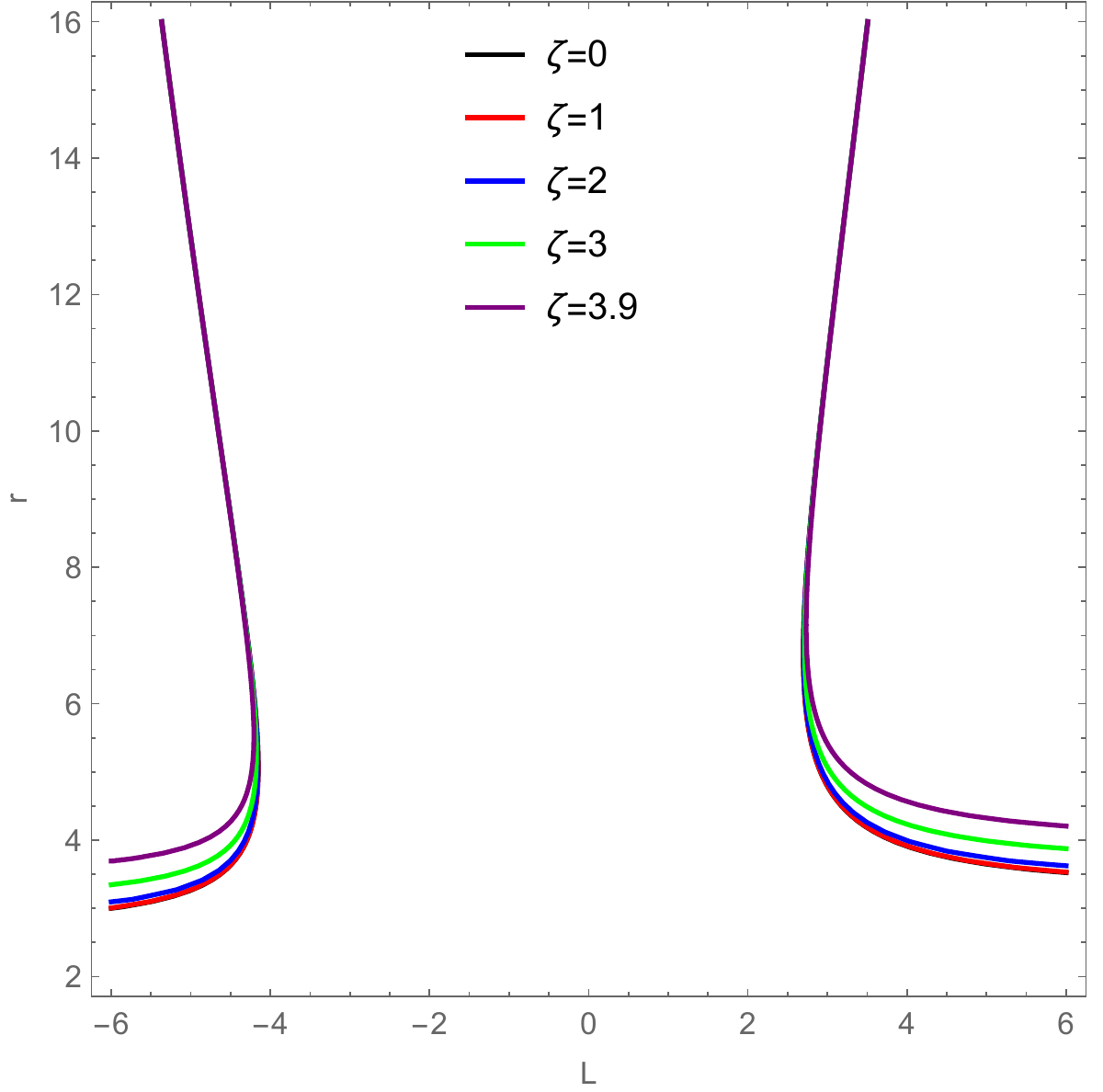}
	\end{subfigure}
	\hfill
	\begin{subfigure}{0.45\textwidth}
		\includegraphics[width=3in, height=5in, keepaspectratio]{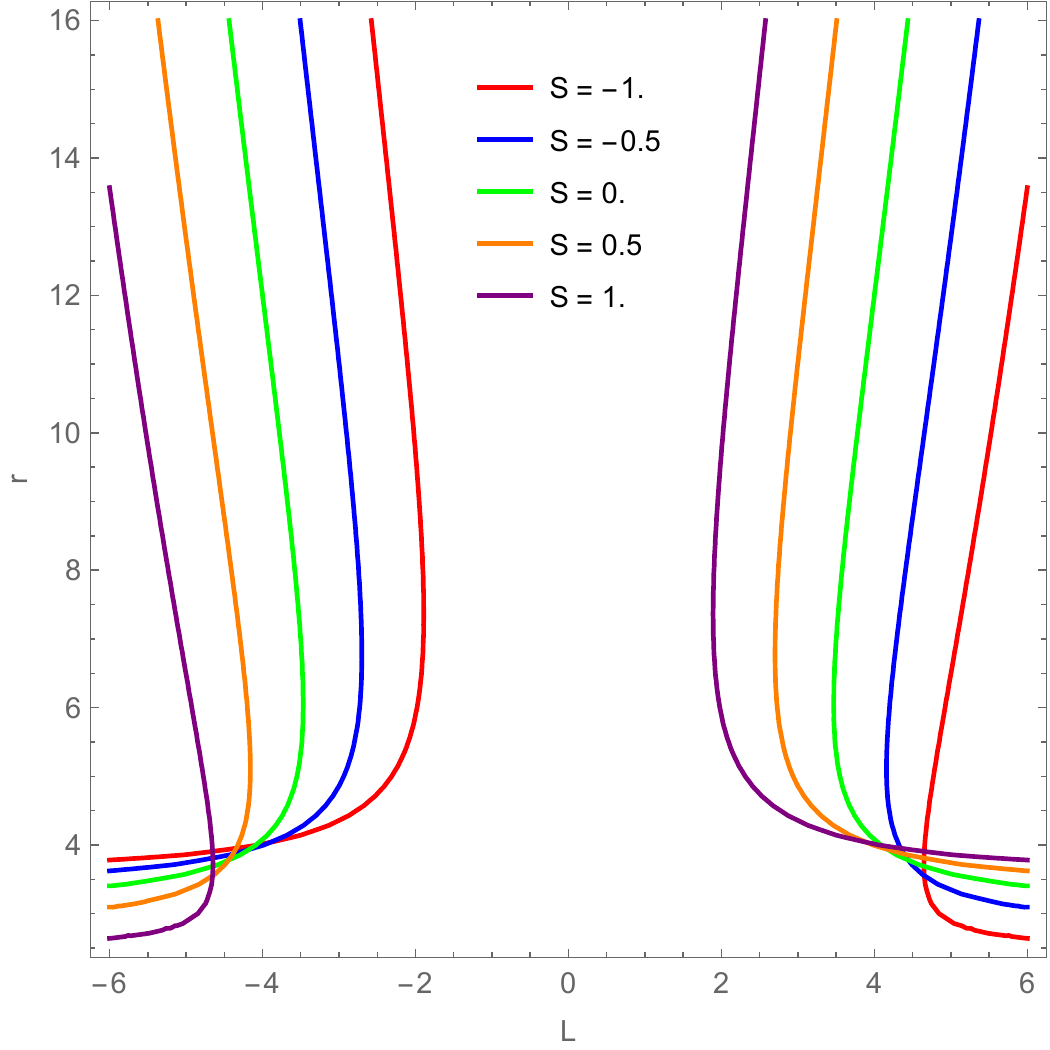}
	\end{subfigure}
	\caption{Behavior of the circular orbit radius $r$ for spinning particles around a quantum-corrected BH with fixed $S=0.5$ (left panel) and fixed $\zeta=2$ (right panel), respectively.}
	\label{fig_radius}
\end{figure*}

The radii of circular orbits (both stable and unstable) as functions of the orbital angular momentum $L$ for different values of the parameters $\zeta$ and $S$ are shown in Fig.~\ref{fig_radius}. The left panel of Fig.~\ref{fig_radius} displays the variation of the circular orbit radius with $L$ at a fixed spin $S=0.5$. The right panel shows the dependence of the circular orbit radius $r$ on $L$ at a fixed $\zeta=2$.

Note that in Fig.~\ref{fig_radius}, there exists a turning point on each curve, corresponding to the minimum absolute value of $L$. The radius curves below this turning point correspond to unstable circular orbits, while those above correspond to stable circular orbits. This point represents ISCO for spinning particles around the quantum-corrected BH. It is evident that when $S$ is fixed, $\zeta$ has a significant influence on unstable circular orbits, while its effect on the radii of stable circular orbits is nearly negligible. In contrast, when $\zeta$ is fixed, variations in $S$ markedly affect the radii of circular orbits. In addition, from the right panel of Fig.~\ref{fig_radius}, we can clearly observe that curves with the same absolute value of $S$ exhibit symmetry about the $L=0$ axis. This symmetry arises because the orbits for test particles with $S>0$ and $L>0$ are identical to those with $S<0$ and $L<0$, and similarly, orbits with $S>0$ and $L<0$ are identical to those with $S<0$ and $L>0$.

\subsection{ISCO around quantum-corrected BHs}

For particles in ISCO, the effective potential for radial motion should satisfy not only Eq.~\eqref{circle} but also the following condition
\begin{equation}
	\begin{split}
		\frac{{\rm d}^2 V_\text{eff}}{{\rm d}r^2} = 0.\label{ISCO}
	\end{split}
\end{equation}
Therefore, using Eqs.~\eqref{circle} and \eqref{ISCO}, we can obtain the ISCO-related physical quantities for different parameter values. Subsequently, in Fig.~\ref{ISCO}, we respectively show the variations of $r_{\rm ISCO}$, $L_{\rm ISCO}$, and $E_{\rm ISCO}$ with the quantum parameter $\zeta$ and spin $S$.
\begin{figure*}[htbp]
	\centering
	\begin{subfigure}{0.33\textwidth}
		\includegraphics[width=2.4in, keepaspectratio]{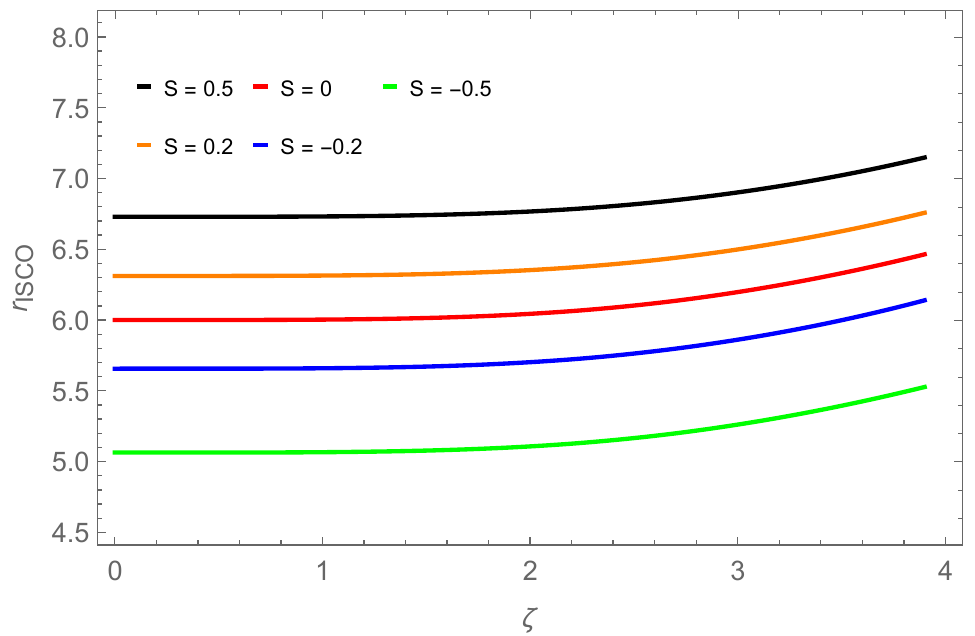}
	\end{subfigure}
	\begin{subfigure}{0.33\textwidth}
		\includegraphics[width=2.4in]{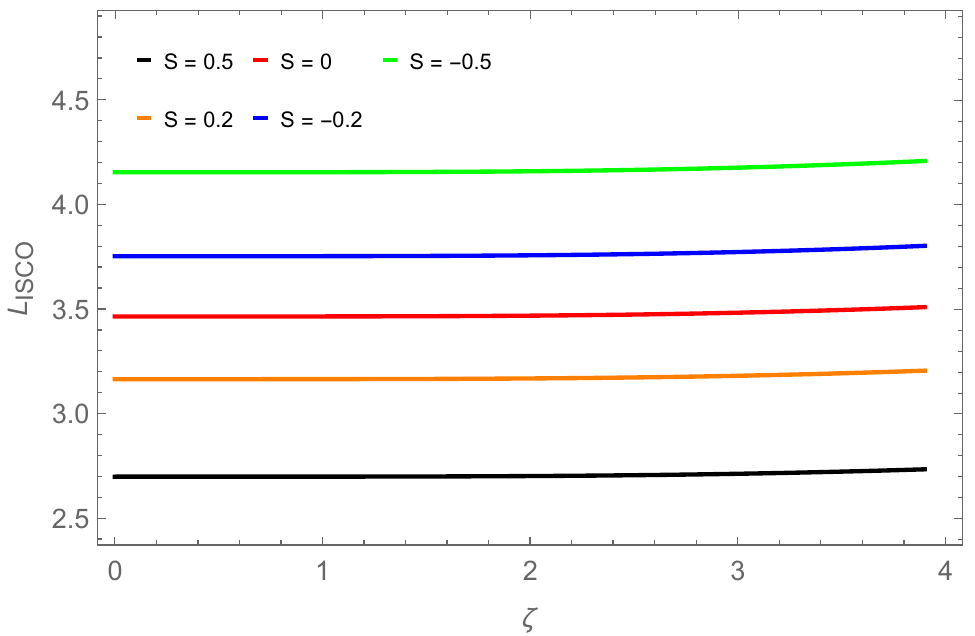}
	\end{subfigure}
	\begin{subfigure}{0.33\textwidth}
		\includegraphics[width=2.4 in,keepaspectratio]{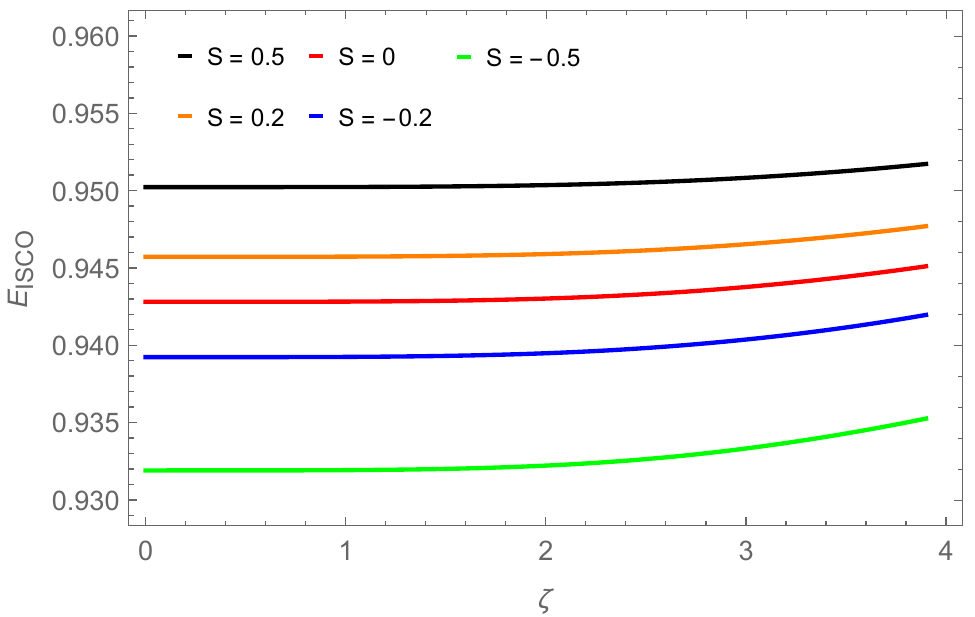}
	\end{subfigure}

	\begin{subfigure}{0.33\textwidth}
		\includegraphics[width=2.4in, keepaspectratio]{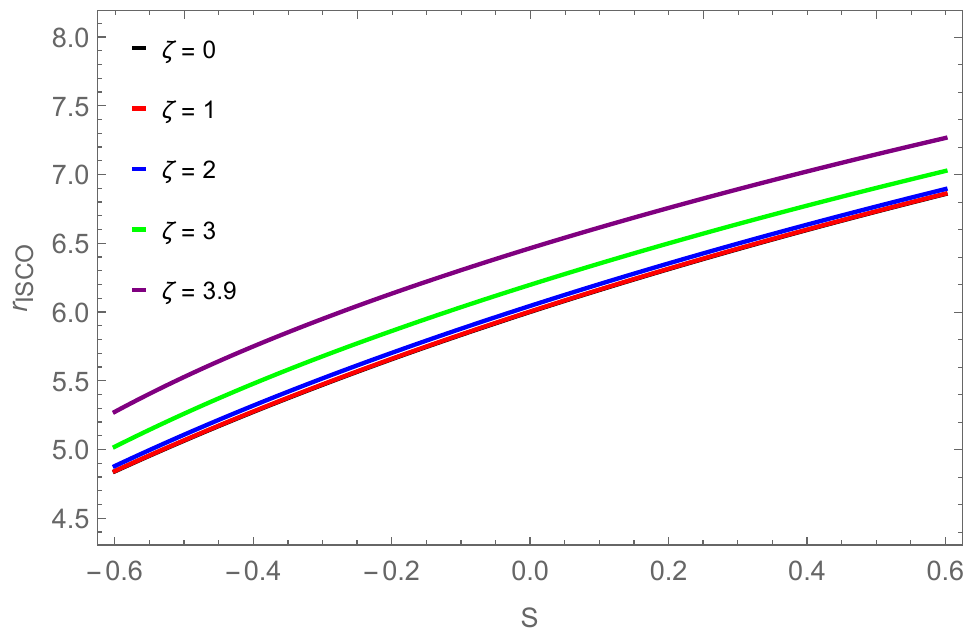}
	\end{subfigure}
	\begin{subfigure}{0.33\textwidth}
		\includegraphics[width=2.4in]{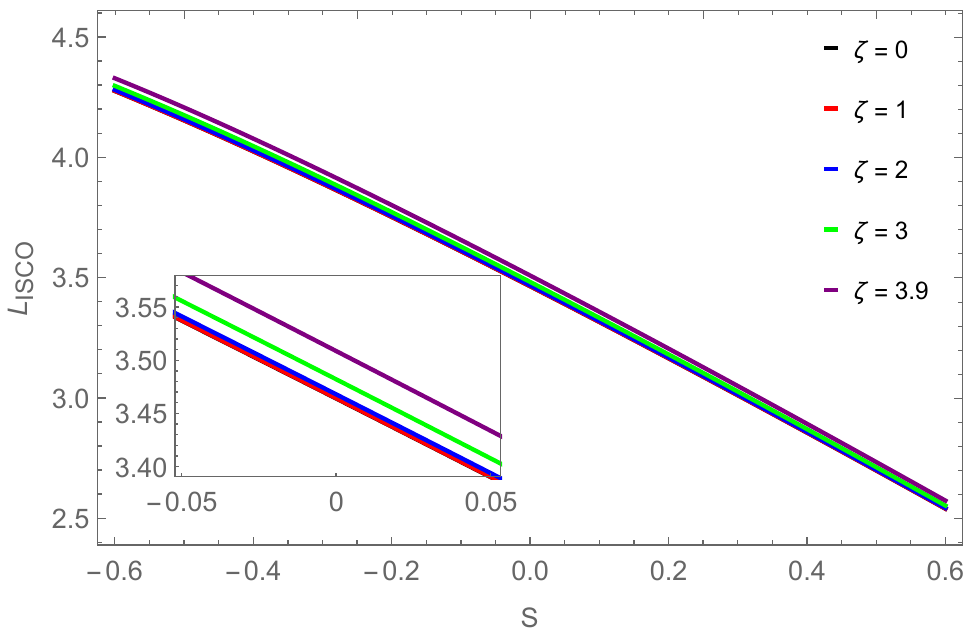}
	\end{subfigure}
	\begin{subfigure}{0.33\textwidth}
		\includegraphics[width=2.4 in,keepaspectratio]{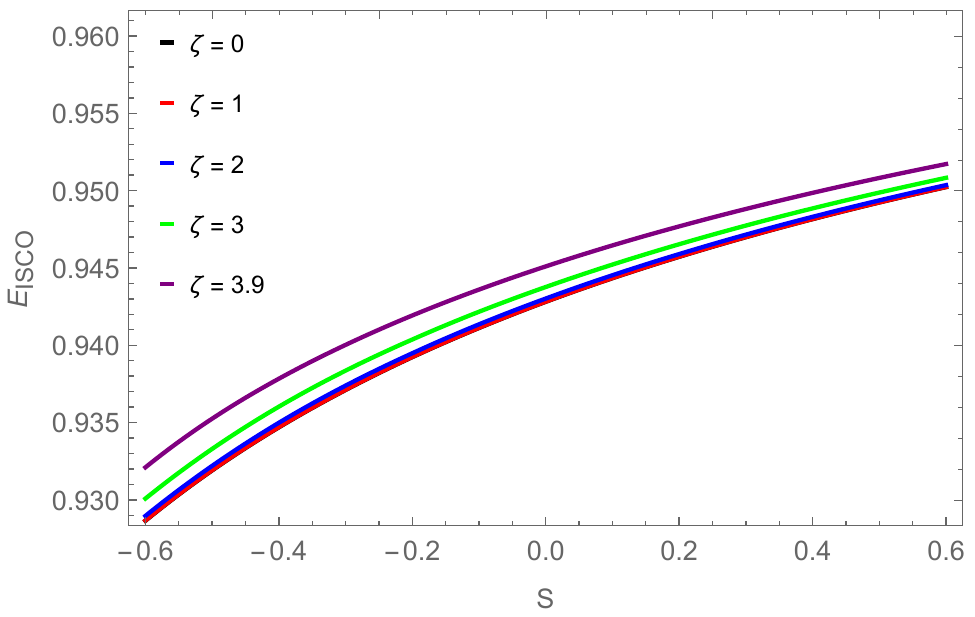}
	\end{subfigure}

	\caption{Variation of ISCO physical quantities ($r_{\rm ISCO}$, $L_{\rm ISCO}$, $E_{\rm ISCO}$) with the quantum parameter $\zeta$ and spin $S$.}
	\label{fig:ISCO}
\end{figure*}

The first row of Fig.~\ref{fig:ISCO} shows the variations of $r_{\rm ISCO}$, $L_{\rm ISCO}$, and $E_{\rm ISCO}$ with the quantum parameter $\zeta$ for different values of $S$. We find that an increase in $\zeta$ consistently leads to increases in $r_{\rm ISCO}$, $L_{\rm ISCO}$, and $E_{\rm ISCO}$, although this effect is weak and not pronounced. The second row reflects the trends of the ISCO physical quantities with $S$ for different values of $\zeta$. It is not difficult to observe that the influence of $S$ on the ISCO physical quantities is more significant than that of $\zeta$.

\subsection{Timelike condition for spinning particles at ISCO}

From Eqs.~\eqref{MPD1}--~\eqref{m}, we obtain the following relation between the 4-velocity and the 4-momentum, which reads~\cite{Tan:2024hzw}
\begin{equation}
	u^a-\frac{P^a}{m}=\frac{S^{ab}S^{de}u^c R_{bcde}}{2m^2}.
\end{equation}
We can see that the 4-velocity and 4-momentum of a spinning particle are not parallel, which may lead to the 4-velocity becoming spacelike--a situation that is unphysical. Therefore, we need to impose a timelike condition constraint on the particle's 4-velocity
\begin{equation}
	\begin{split}
	u^a u_a &=g_{tt}(\frac{dt}{d\tau})^2+g_{rr}(\frac{dr}{d\tau})^2+g_{\phi \phi}(\frac{d\phi}{d\tau})^2 \\
	&=g_{tt}+g_{rr}(u^r)^2+g_{\phi \phi}(u^{\phi})^2 \; <\; 0.
	\end{split}
\end{equation}
Here we have adopted the previous assumption $u^t = 1$. Taking a spinning particle at the ISCO around a quantum-corrected BH as an example, we derive the constraint between the spin $S$ and the quantum parameter $\zeta$ when the particle satisfies the timelike condition at the ISCO. To this end, we first use the effective potential $V_{\rm eff}$ to determine $r_{\text{ISCO}}$, $L_{\text{ISCO}}$, and $E_{\text{ISCO}}$ for given $S$ and $\zeta$. We can then solve for the 4-velocity of the particle using Eqs.~\eqref{vr} and \eqref{vphi}, and subsequently check whether the timelike condition is satisfied.

\begin{figure}[htbp]
	\centering
	\begin{subfigure}{0.45\textwidth}
		\includegraphics[width=3.2in, height=5in, keepaspectratio]{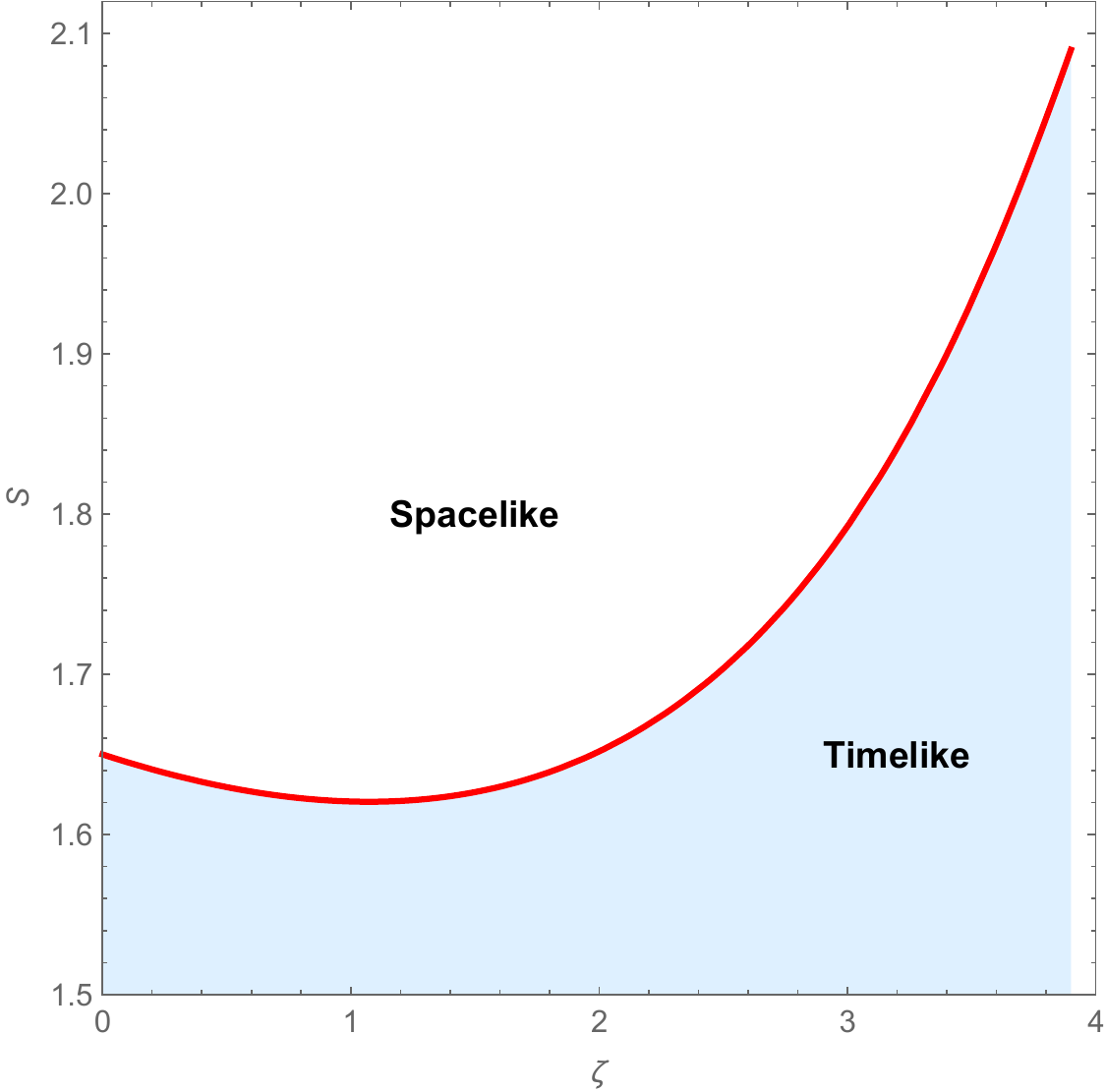}

	\end{subfigure}
	\caption{Constraints on $S$ and $\zeta$ from the timelike condition for spinning particles at ISCO. The red curve represents $u^a u_a = 0$. The blue region indicates $u^a u_a < 0$, and the blank area corresponds to $u^a u_a > 0$.}
	\label{fig_time}
\end{figure}
In Fig.~\ref{fig_time}, we present the constraints on $S$ and $\zeta$ derived from the timelike condition for spinning particles at the ISCO. The blue region represents the timelike region for the 4-velocity of the spinning particle, while the blank area indicates the spacelike region. The red curve denotes the critical case where $u^a u_a = 0$. It can be clearly observed that as $\zeta$ increases, the critical value of $S$ first decreases slowly and then gradually increases.

\section{Trajectories of Spinning Particles in Quantum-Corrected Spacetime}\label{seciton4}

In the previous section, we obtained the 4-velocity of spinning particles in this quantum-corrected spacetime. Here, we further investigate the trajectories of spinning particles in the equatorial plane around this quantum-corrected BH. The trajectory equations for spinning particles can be derived from Eqs.~\eqref{vr} and \eqref{vphi}, namely
\begin{eqnarray}
	\frac{u^r}{u^{\phi}}=\frac{{\rm d} r}{{\rm d}\phi}=\frac{P^r-\frac{S}{2F}R_{\phi t\mu\nu}S^{\mu\nu}}{P^\phi+\frac{S}{2F}R_{rt\mu\nu}S^{\mu\nu}}.\label{trajectory}
\end{eqnarray}
Thus, by specifying the initial conditions of the spinning particle, we can solve the trajectory equations and plot the particle's trajectory in a two-dimensional plane. When setting the initial conditions, we consider the particle's energy to lie between that of the typical unstable circular orbit and the stable circular orbit. In this case, the particle will oscillate between the periastron and apastron, forming a bound orbit.

\begin{figure*}[htbp]
	\centering
	\begin{subfigure}{0.33\textwidth}
		\includegraphics[height=4cm, keepaspectratio]{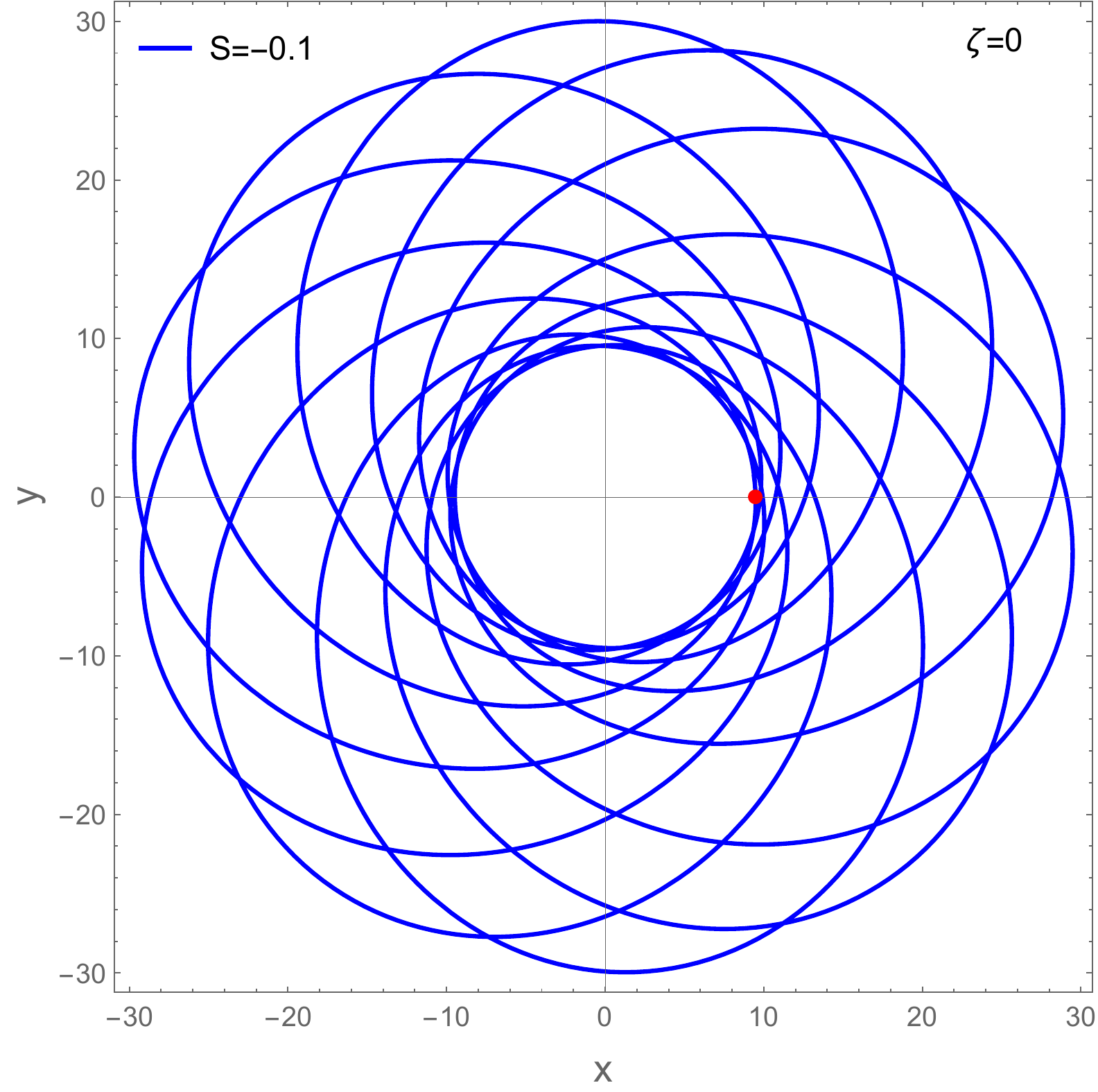}
		\caption{$S=-0.1$ and $\zeta=0$}
	\end{subfigure}
	\begin{subfigure}{0.33\textwidth}
		\includegraphics[height=4cm]{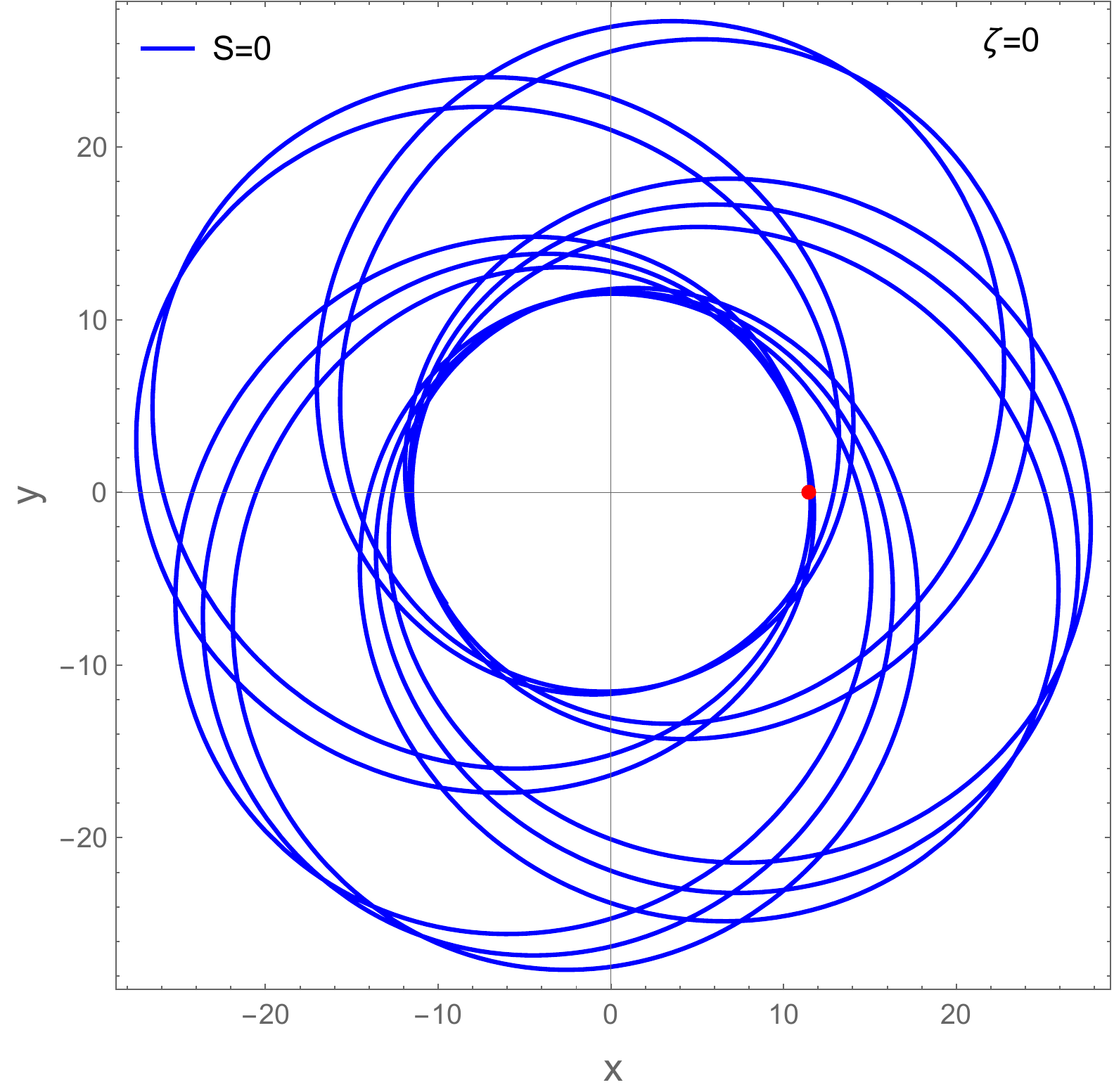}
		\caption{$S=0$ and $\zeta=0$}
	\end{subfigure}
	\begin{subfigure}{0.33\textwidth}
		\includegraphics[height=4cm,keepaspectratio]{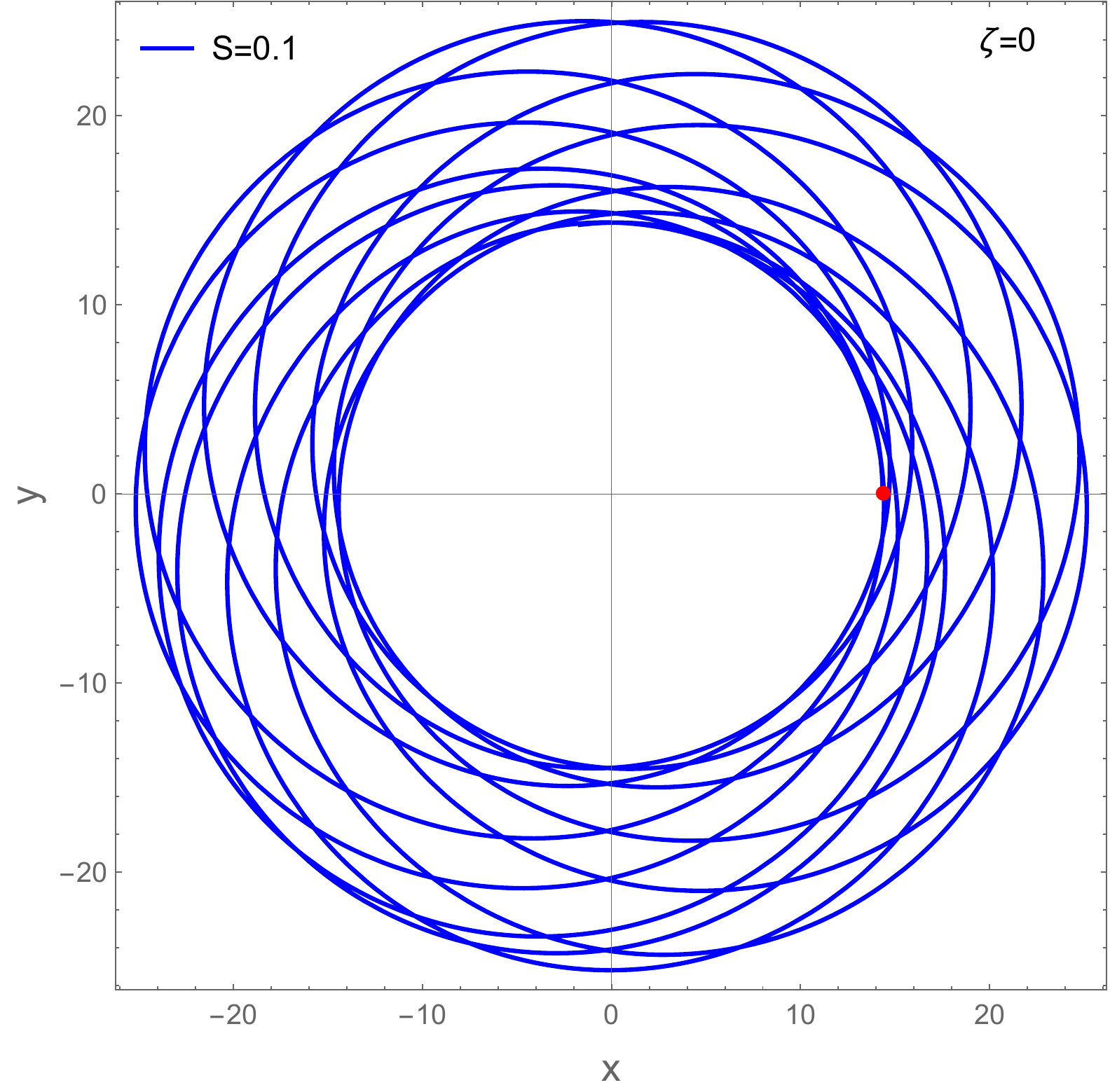}
		\caption{$S=0.1$ and $\zeta=0$}
	\end{subfigure}

	\begin{subfigure}{0.33\textwidth}
		\includegraphics[height=4cm, keepaspectratio]{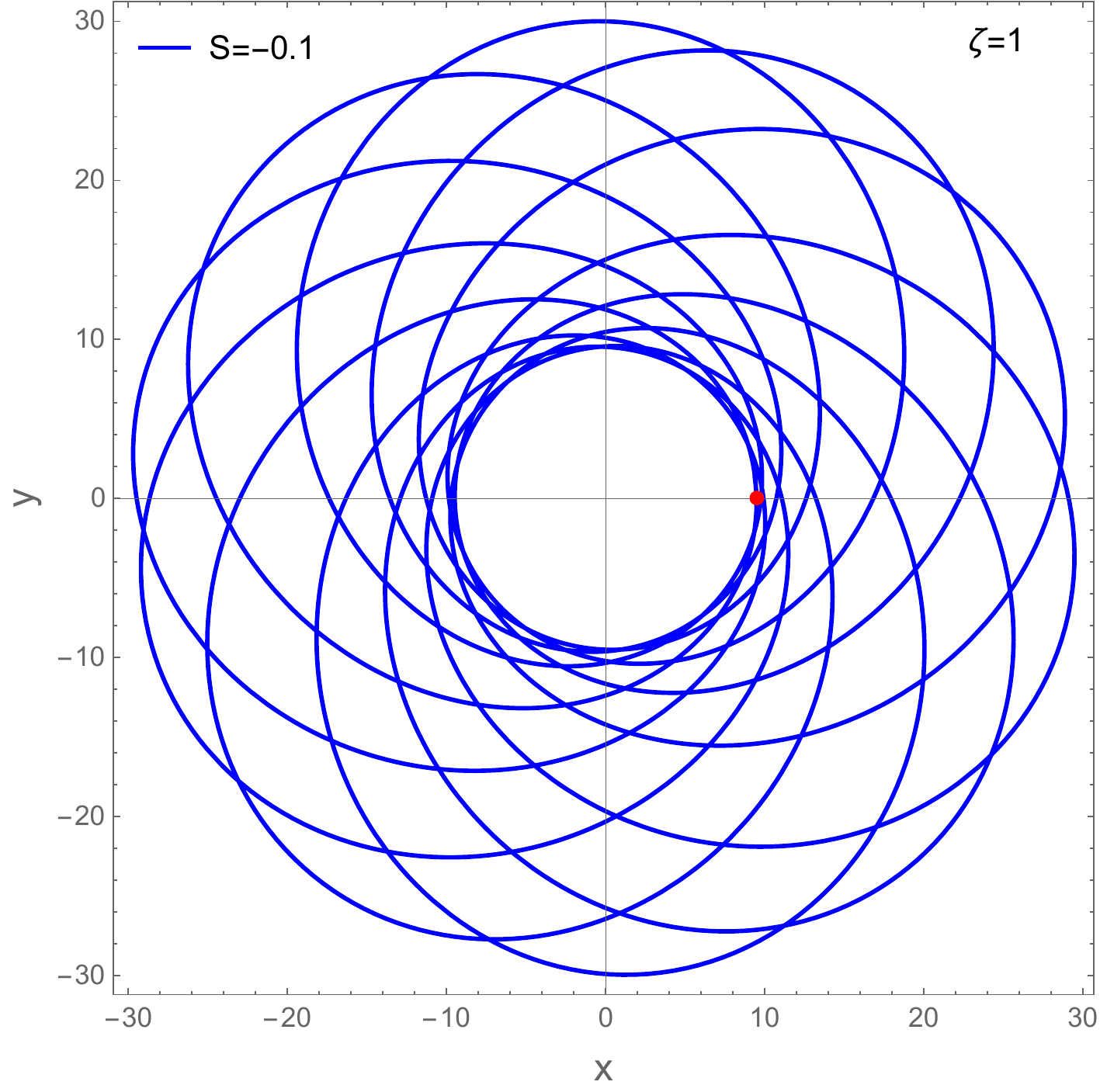}
		\caption{$S=-0.1$ and $\zeta=1$}
	\end{subfigure}
	\begin{subfigure}{0.33\textwidth}
		\includegraphics[height=4cm]{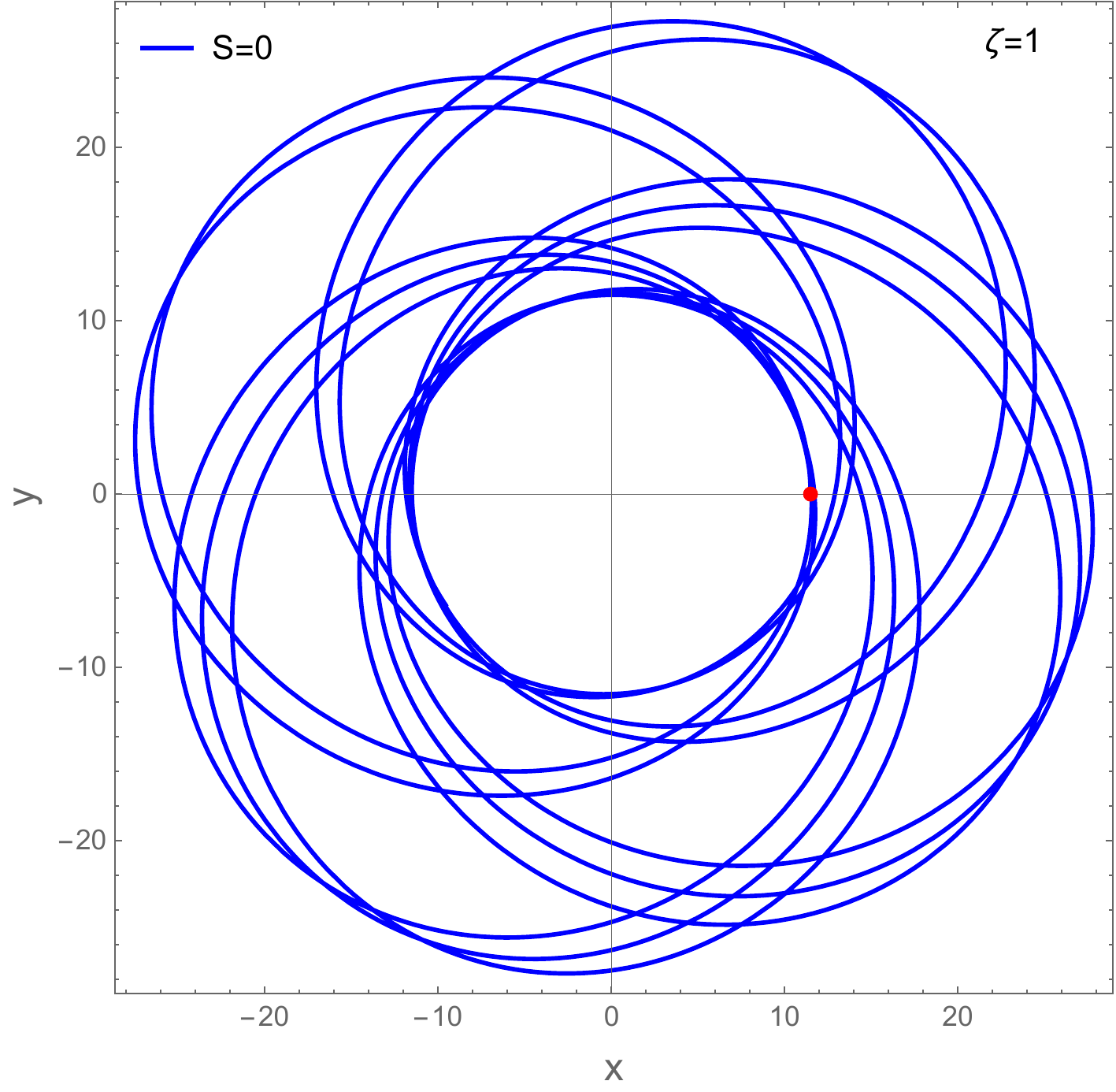}
		\caption{$S=0$ and $\zeta=1$}
	\end{subfigure}
	\begin{subfigure}{0.33\textwidth}
		\includegraphics[height=4cm,keepaspectratio]{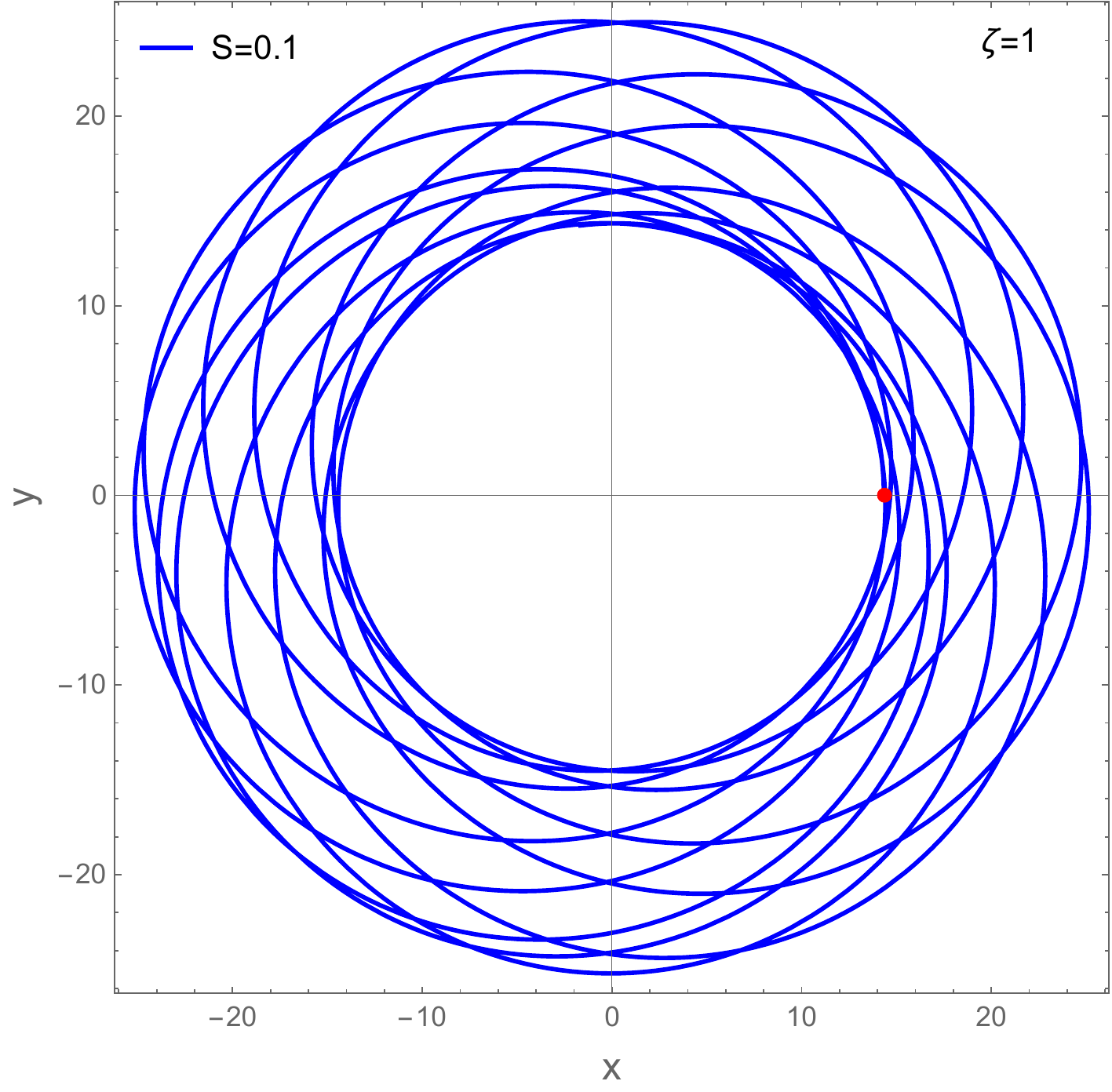}
		\caption{$S=0.1$ and $\zeta=1$}
	\end{subfigure}

	\begin{subfigure}{0.33\textwidth}
		\includegraphics[height=4cm, keepaspectratio]{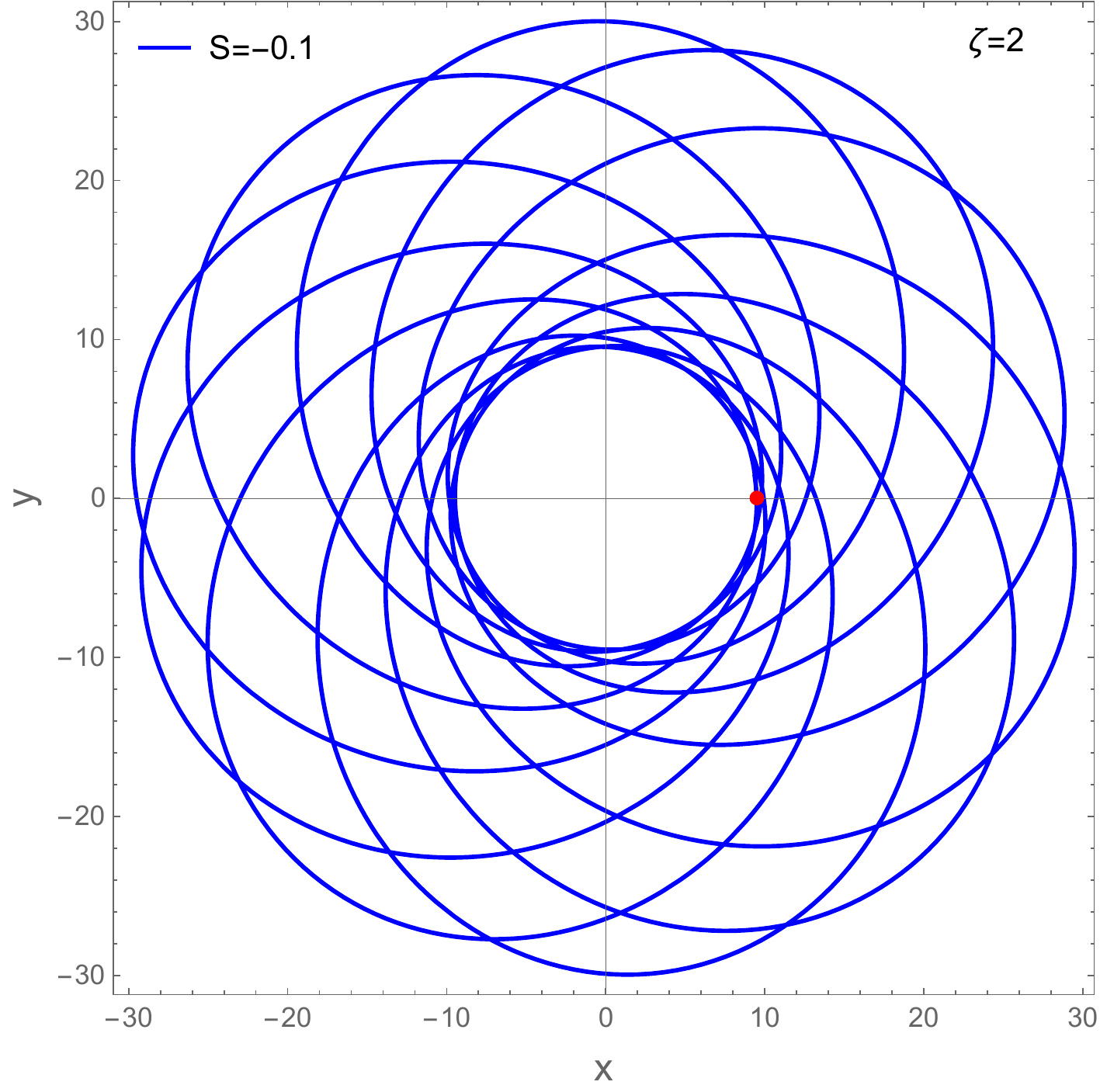}
		\caption{$S=-0.1$ and $\zeta=2$}
	\end{subfigure}
	\begin{subfigure}{0.33\textwidth}
		\includegraphics[height=4cm]{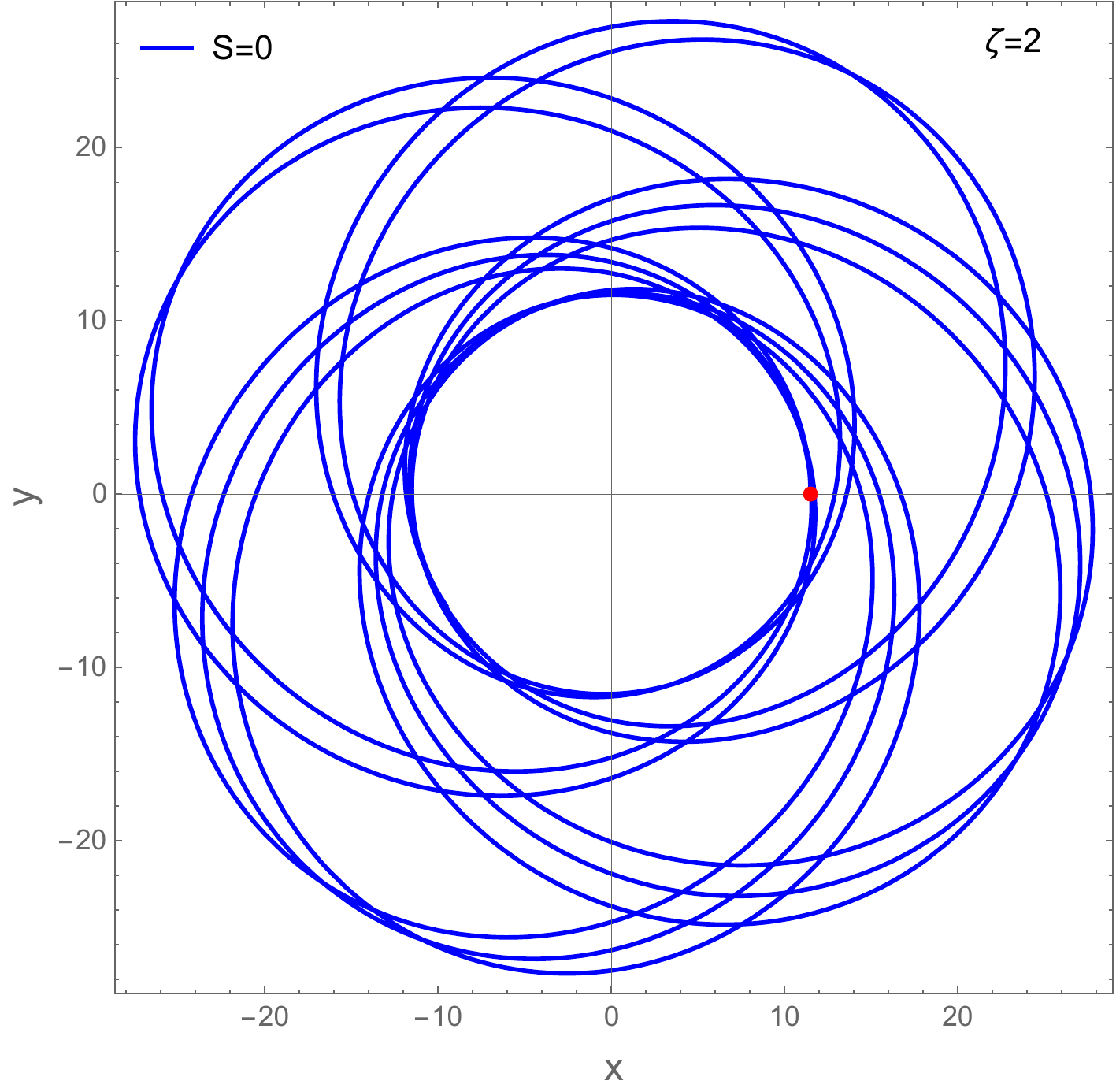}
		\caption{$S=0$ and $\zeta=2$}
	\end{subfigure}
	\begin{subfigure}{0.33\textwidth}
		\includegraphics[height=4cm,keepaspectratio]{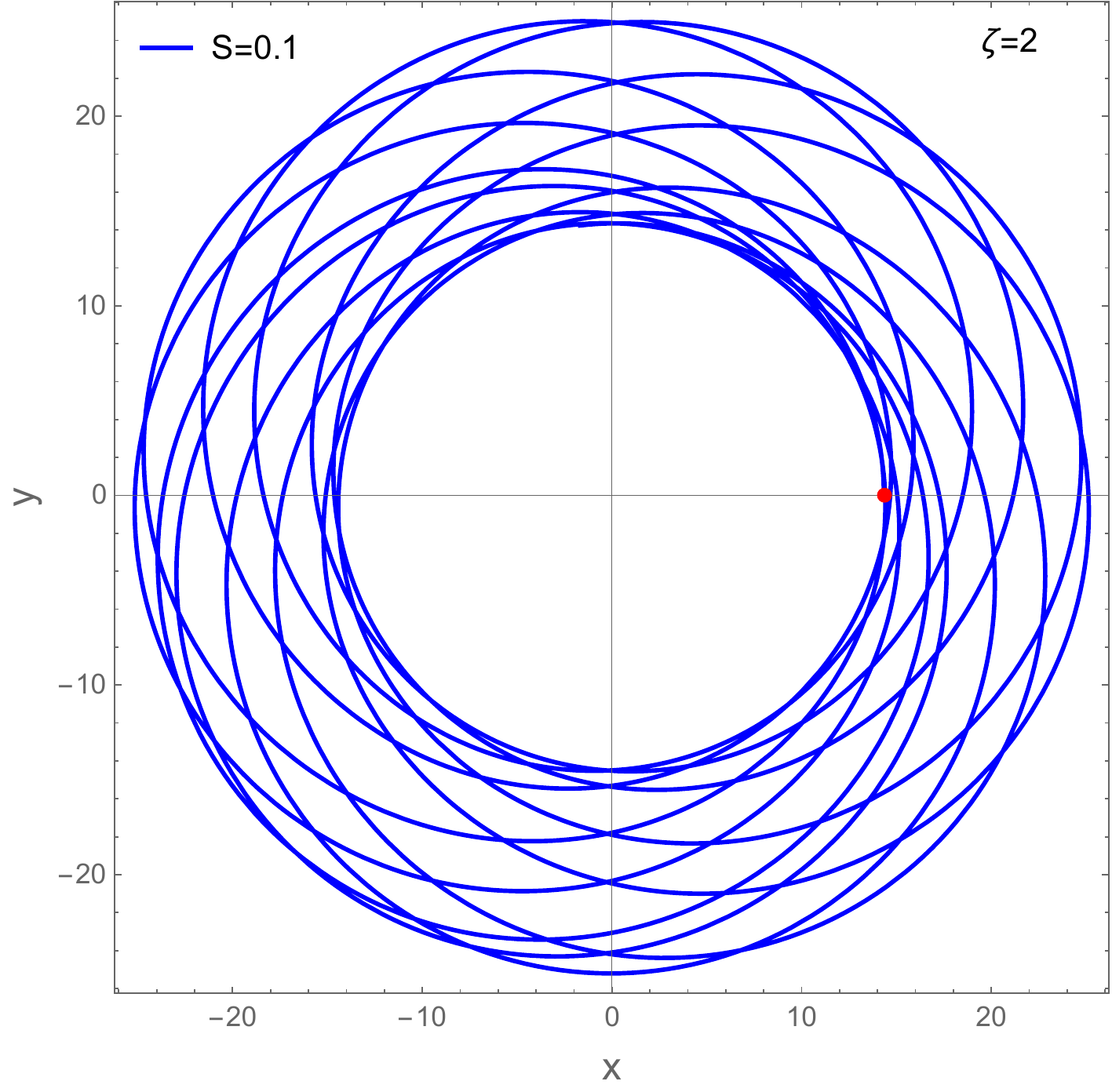}
		\caption{$S=0.1$ and $\zeta=2$}
	\end{subfigure}
	\begin{subfigure}{0.33\textwidth}
		\includegraphics[height=4cm, keepaspectratio]{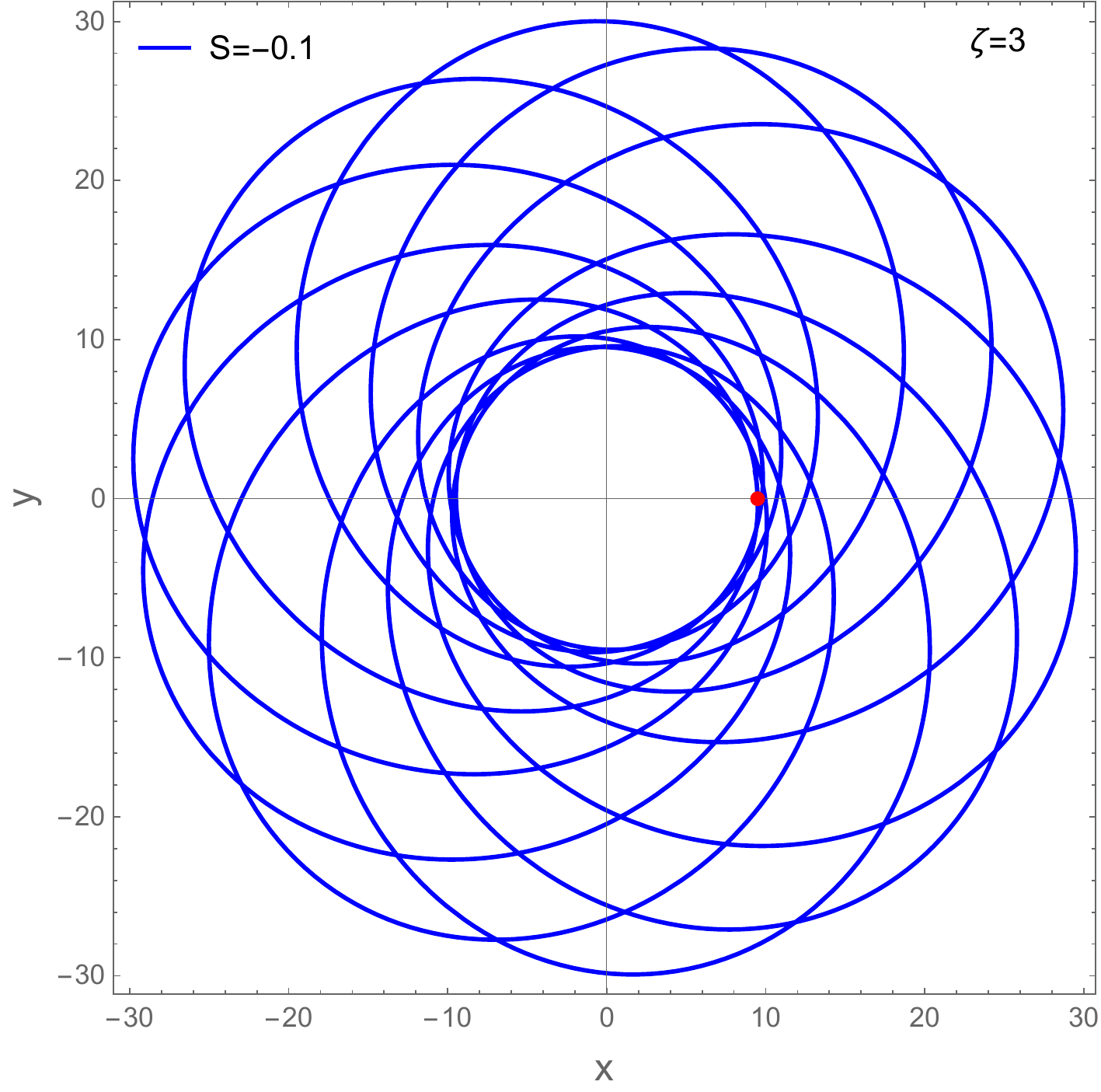}
		\caption{$S=-0.1$ and $\zeta=3$}
	\end{subfigure}
	\begin{subfigure}{0.33\textwidth}
		\includegraphics[height=4cm]{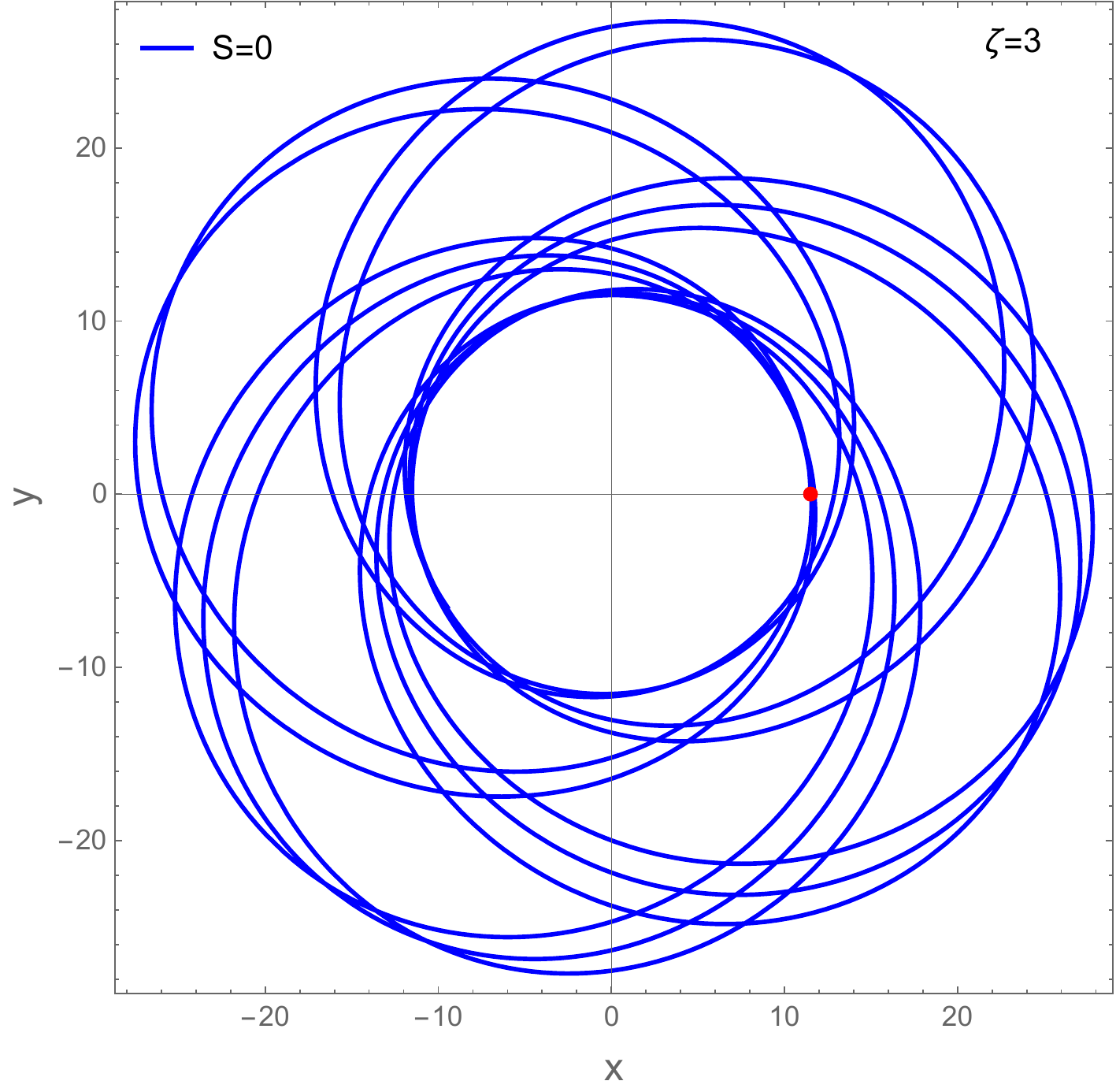}
		\caption{$S=0$ and $\zeta=3$}
	\end{subfigure}
	\begin{subfigure}{0.33\textwidth}
		\includegraphics[height=4cm,keepaspectratio]{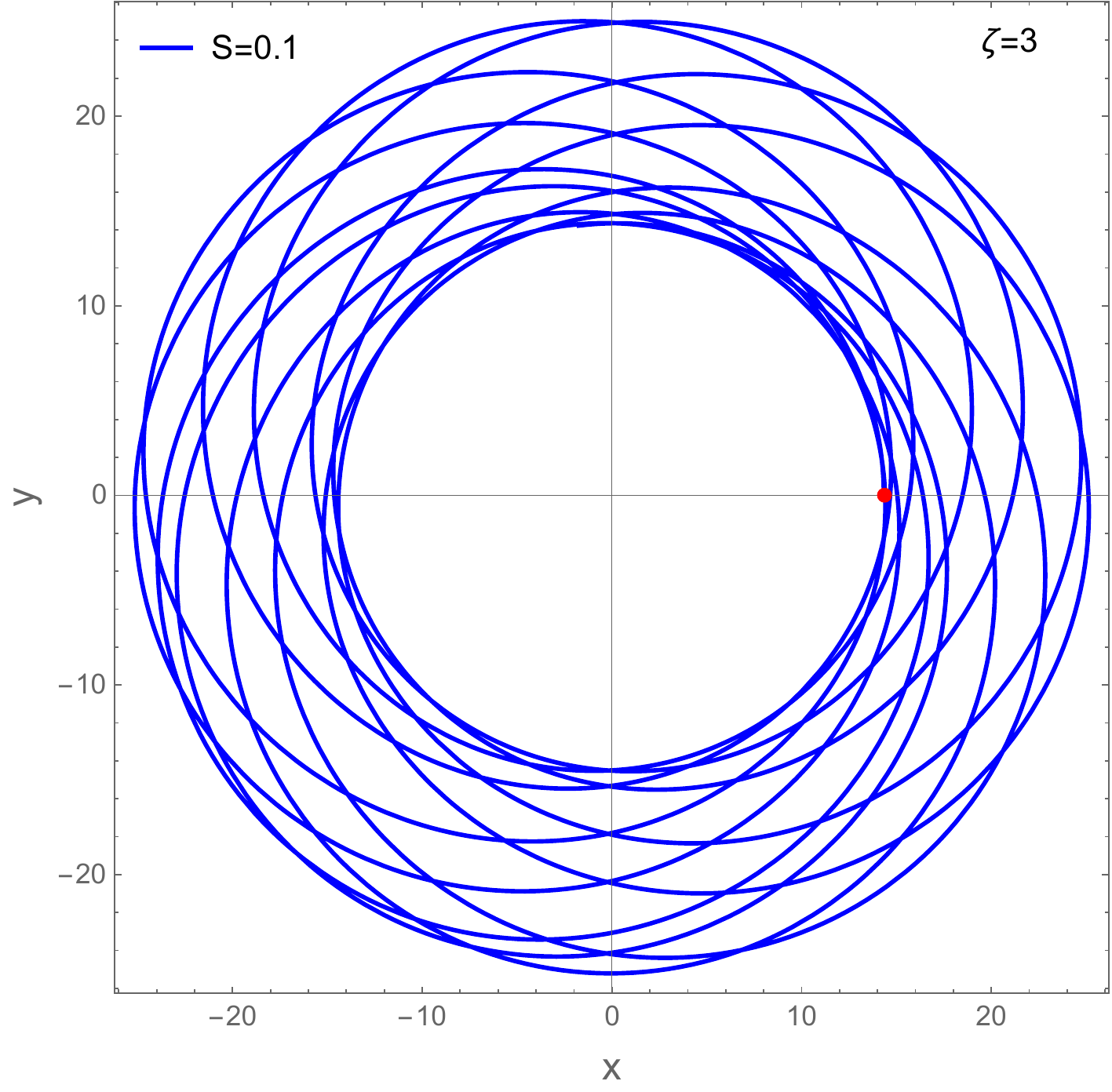}
		\caption{$S=0.1$ and $\zeta=3$}
	\end{subfigure}
	\begin{subfigure}{0.33\textwidth}
		\includegraphics[height=4cm, keepaspectratio]{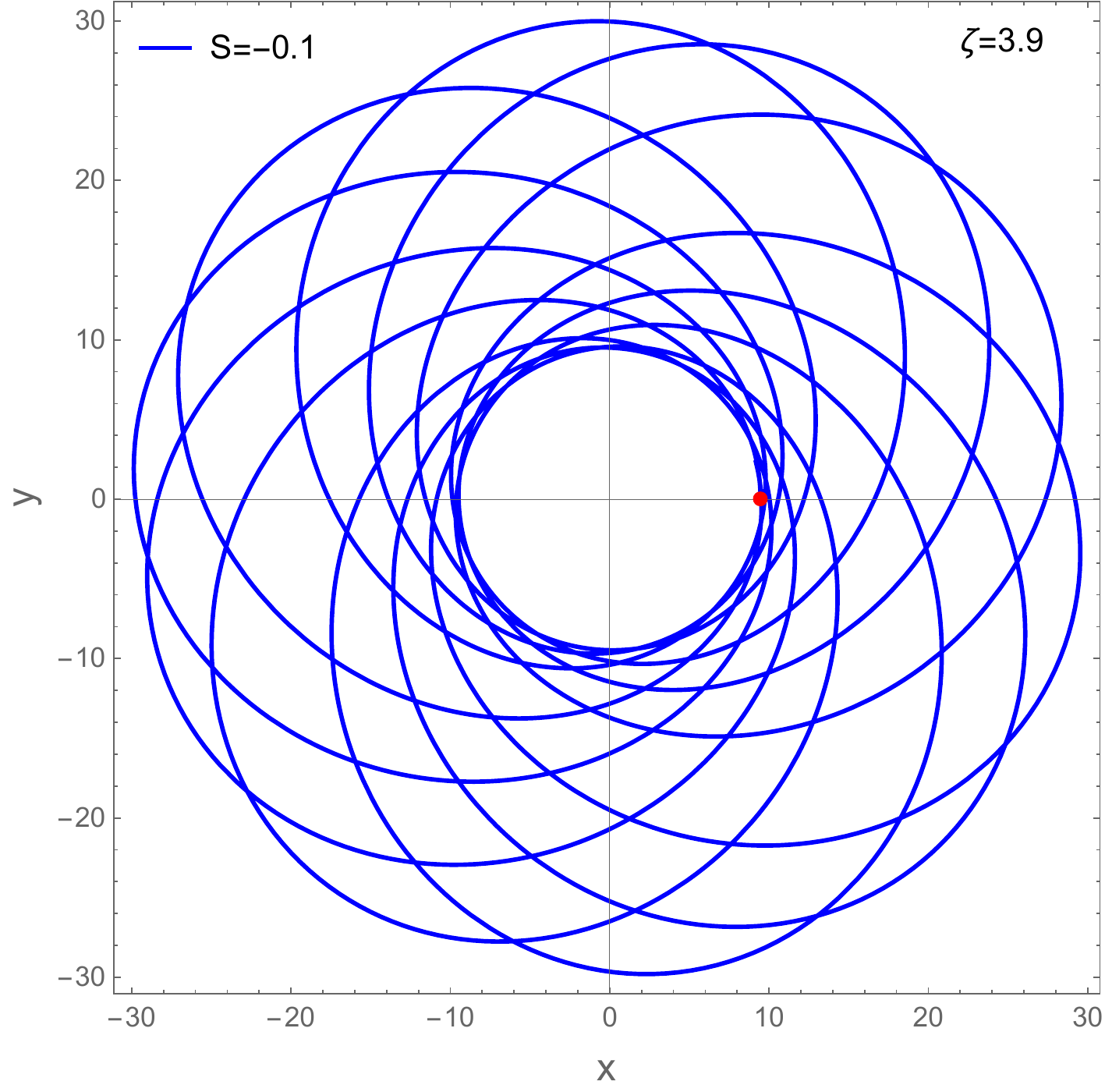}
		\caption{$S=-0.1$ and $\zeta=3.9$}
	\end{subfigure}
	\begin{subfigure}{0.33\textwidth}
		\includegraphics[height=4cm]{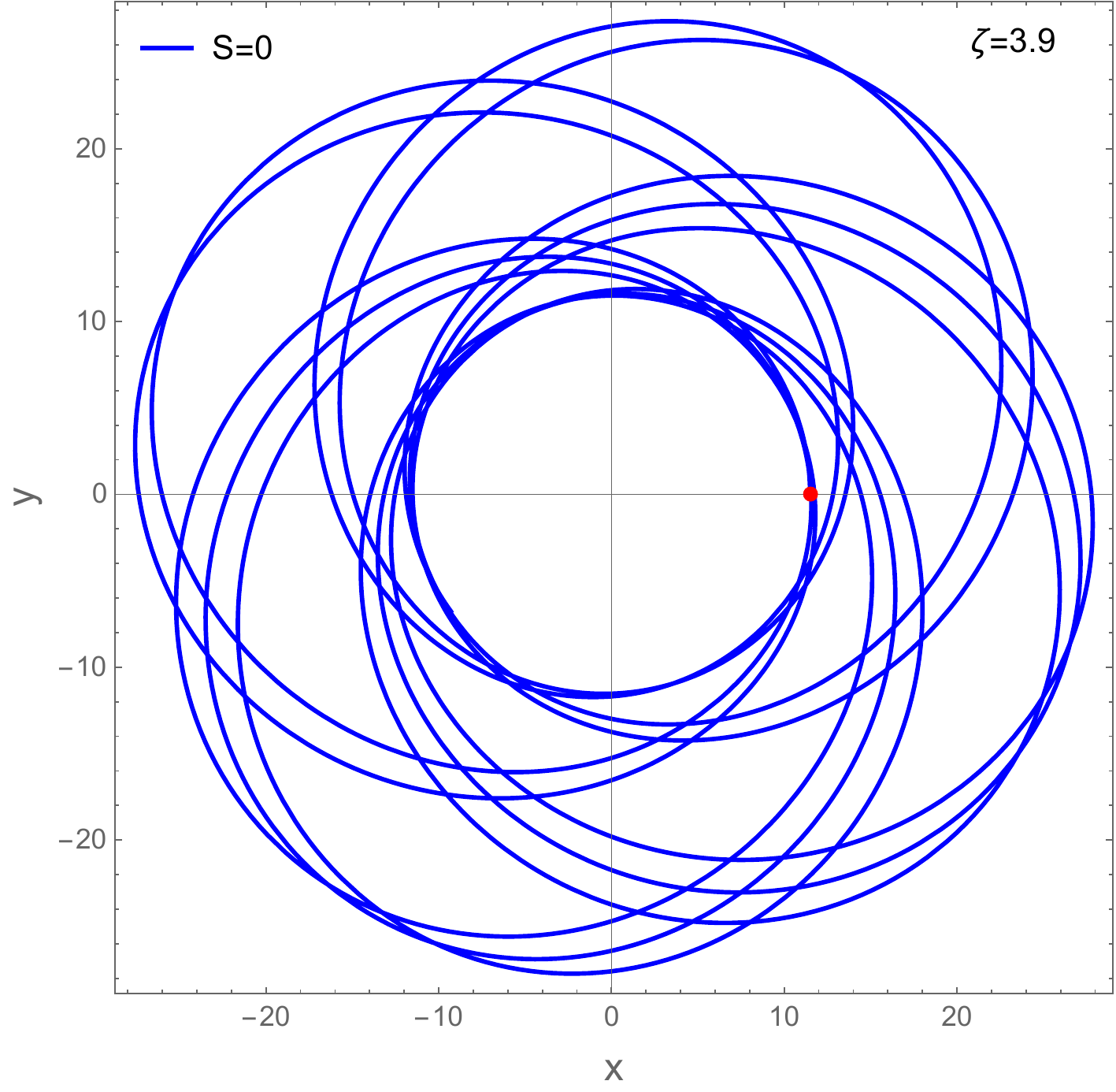}
		\caption{$S=0$ and $\zeta=3.9$}
	\end{subfigure}
	\begin{subfigure}{0.33\textwidth}
		\includegraphics[height=4cm,keepaspectratio]{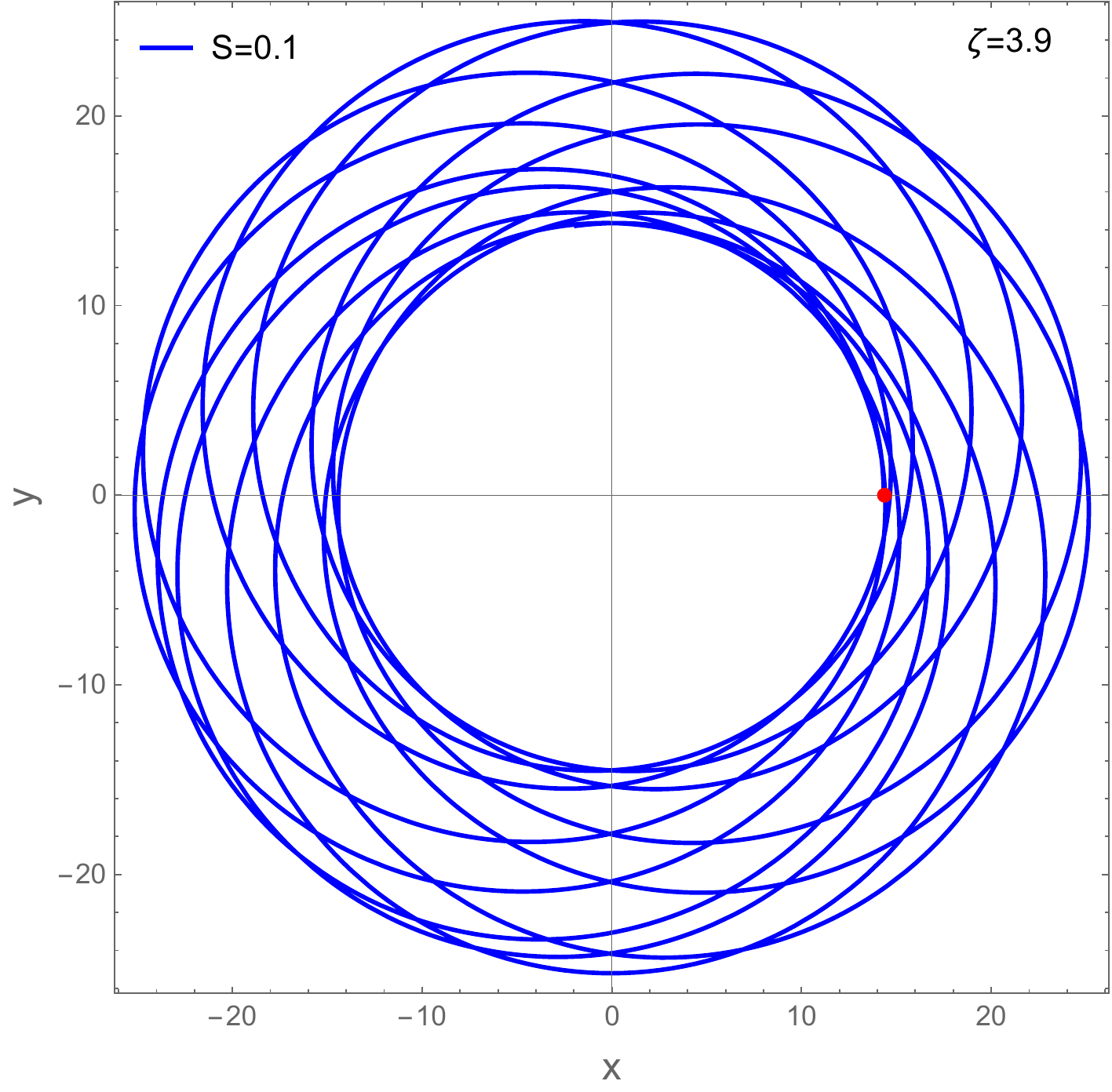}
		\caption{$S=0.1$ and $\zeta=3.9$}
	\end{subfigure}
	\caption{Trajectories of spinning particles with $E=0.976$ and $L=4.5$ in the equatorial plane around BH-III for different values of spin $S$ and quantum parameter $\zeta$.}
	\label{fig:guiji}
\end{figure*}
Here we consider the trajectories of spinning particles with energy $E=0.976$ and orbital angular momentum $L=4.5$ for spin values $S=-0.1, 0, 0.1$ under different values of the quantum parameter $\zeta$, as shown in Fig.~\ref{fig:guiji}. We define one complete cycle as the motion of a spinning particle from the periastron to the apastron and back to the periastron. The trajectories of spinning particles shown here are all obtained after the same number of such cycles. In the panels of Fig.~\ref{fig:guiji}, the red dots indicate the starting points of the particles from the periastron. It can be clearly observed that for the same $\zeta$, different values of $S$ produce significantly distinct trajectories. In contrast, when $S$ is small and fixed, the influence of $\zeta$ on the trajectories is relatively weak, making them almost indistinguishable from the Schwarzschild case ($\zeta=0$).

\begin{figure*}[htbp]
	\centering

	\begin{subfigure}{0.33\textwidth}
		\includegraphics[height=5cm, keepaspectratio]{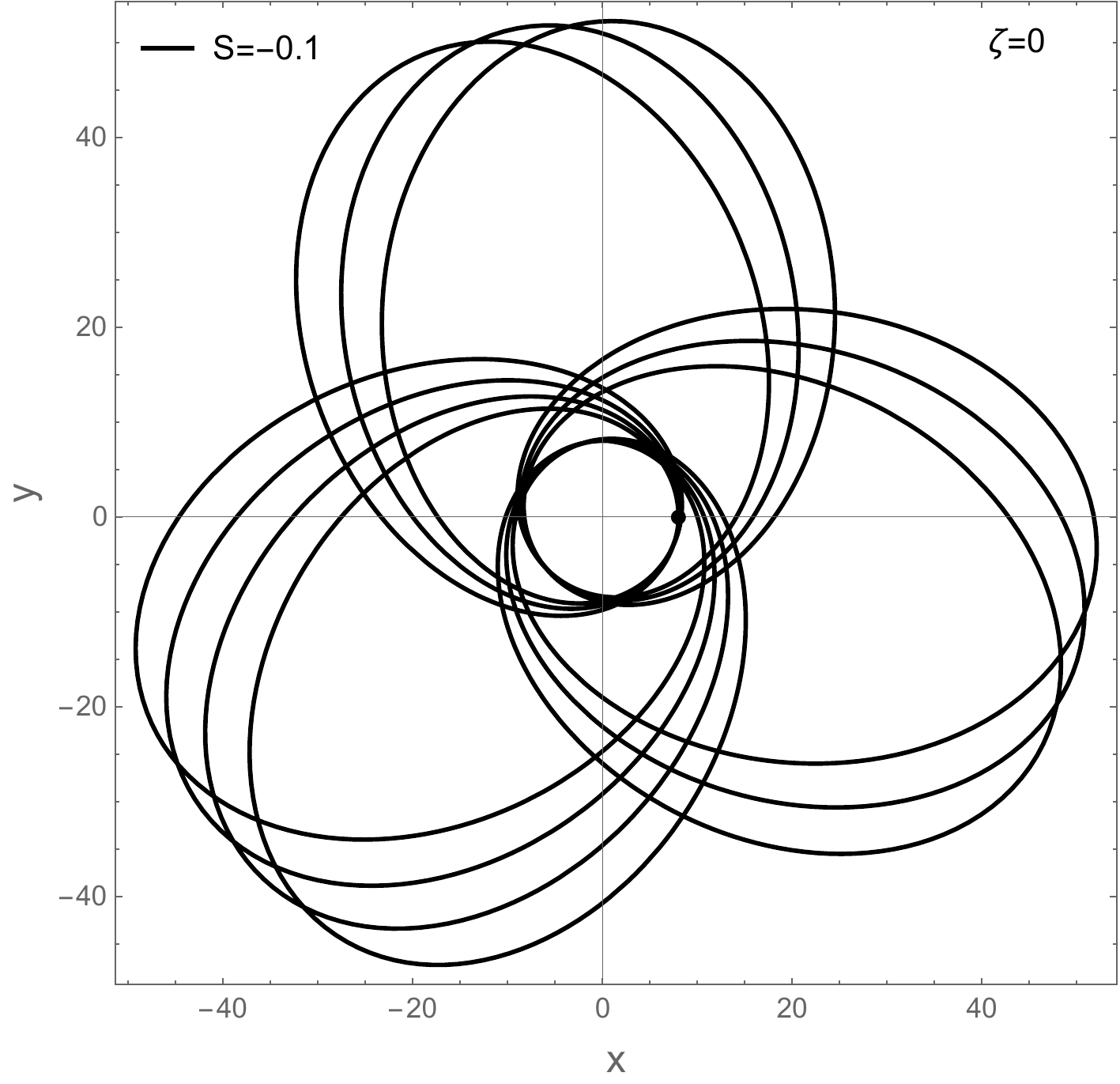}
		\caption{Schwarzschild BH}
	\end{subfigure}
	\begin{subfigure}{0.33\textwidth}
		\includegraphics[height=5cm]{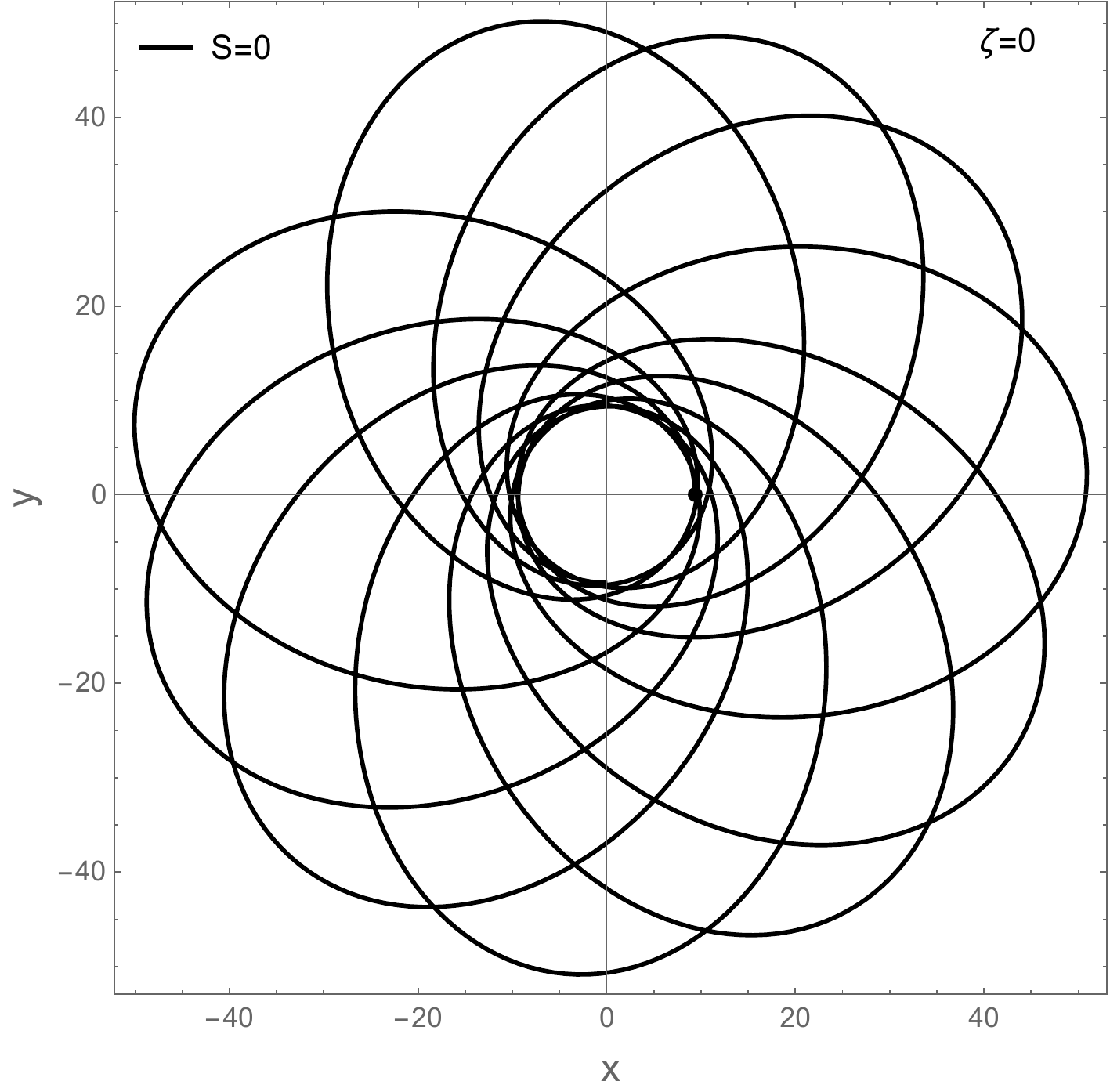}
		\caption{Schwarzschild BH}
	\end{subfigure}
	\begin{subfigure}{0.33\textwidth}
		\includegraphics[height=5cm,keepaspectratio]{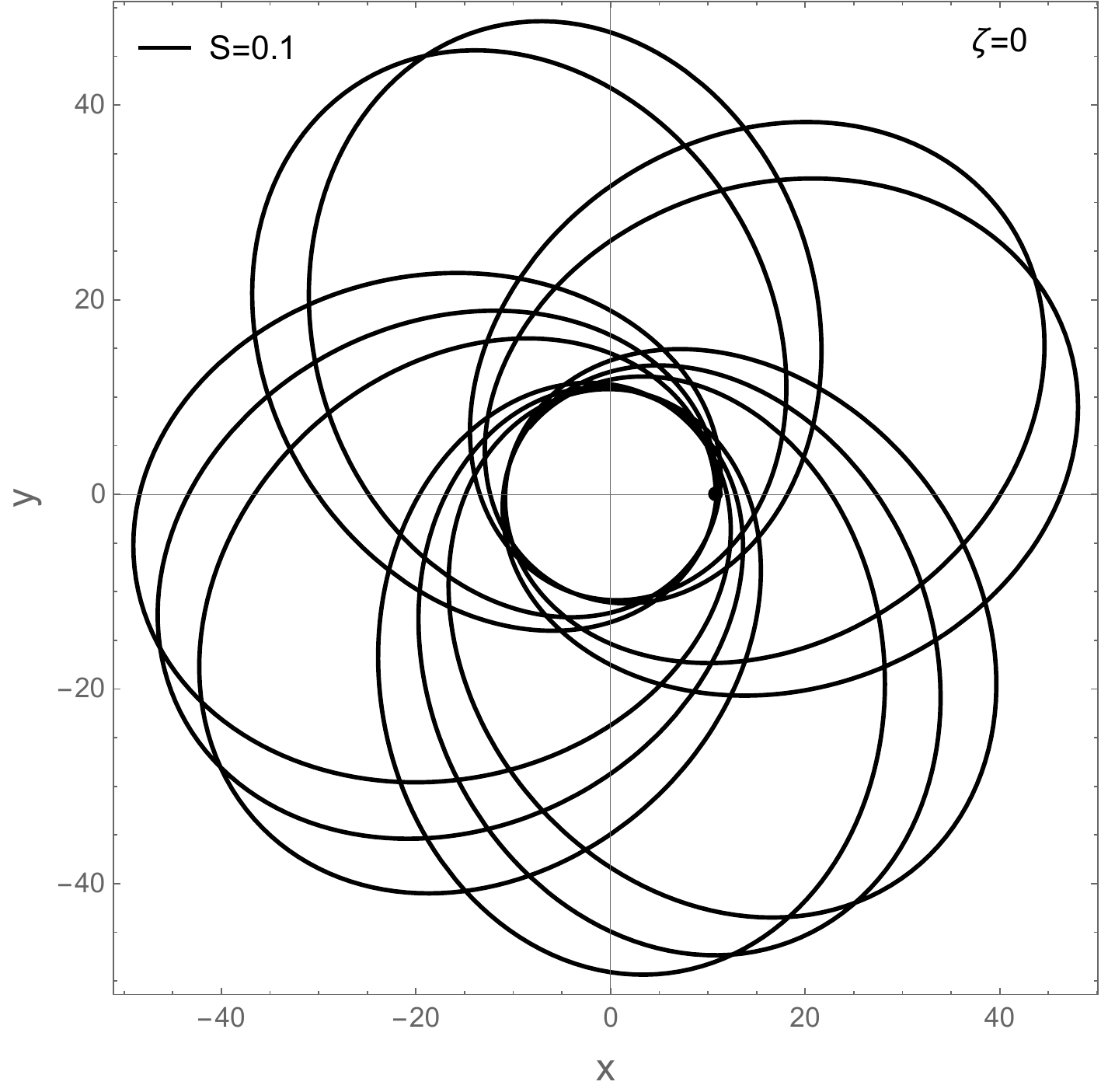}
		\caption{Schwarzschild BH}
	\end{subfigure}

	\begin{subfigure}{0.33\textwidth}
		\includegraphics[height=5cm, keepaspectratio]{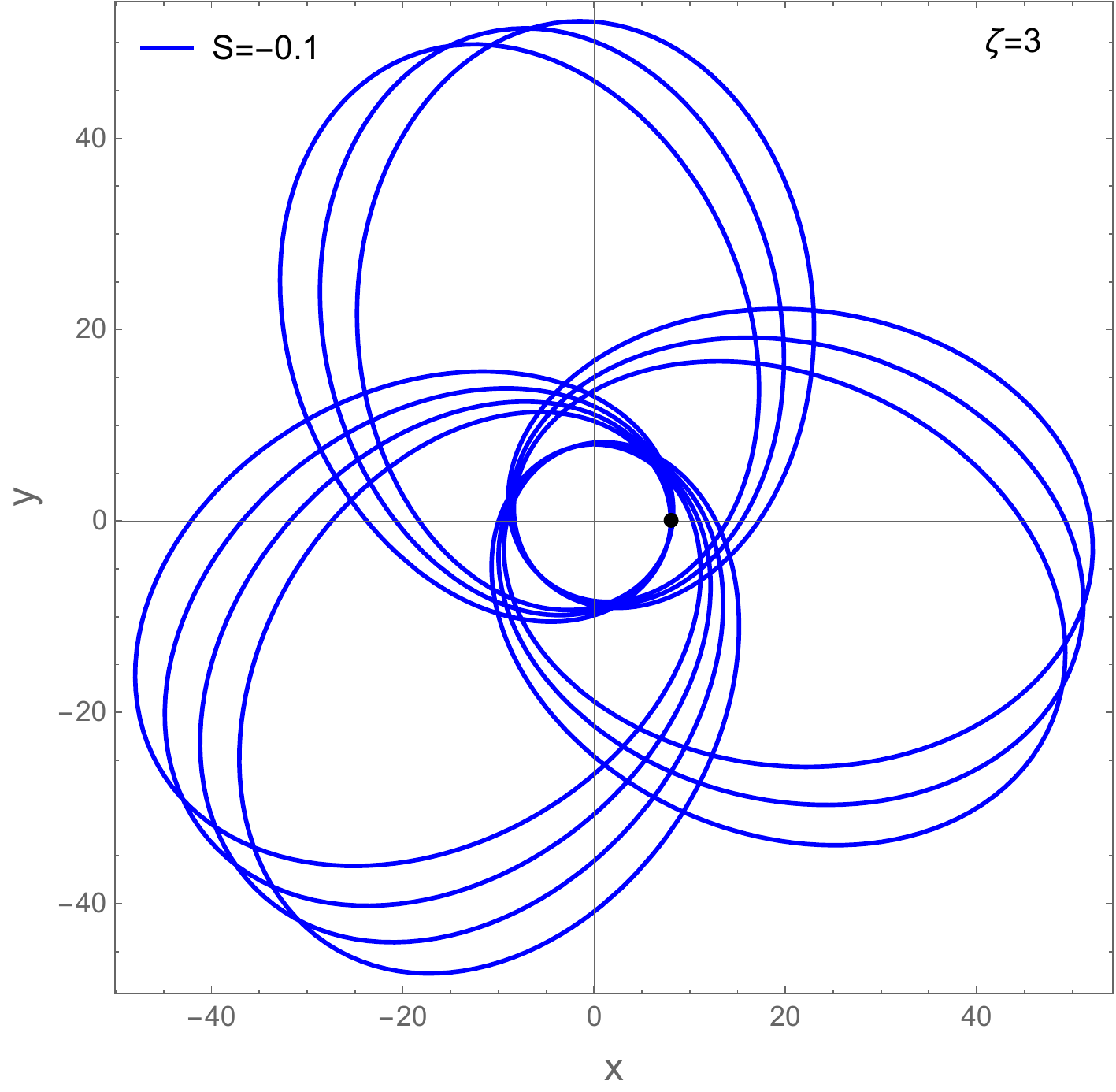}
		\caption{BH-III}
	\end{subfigure}
	\begin{subfigure}{0.33\textwidth}
		\includegraphics[height=5cm]{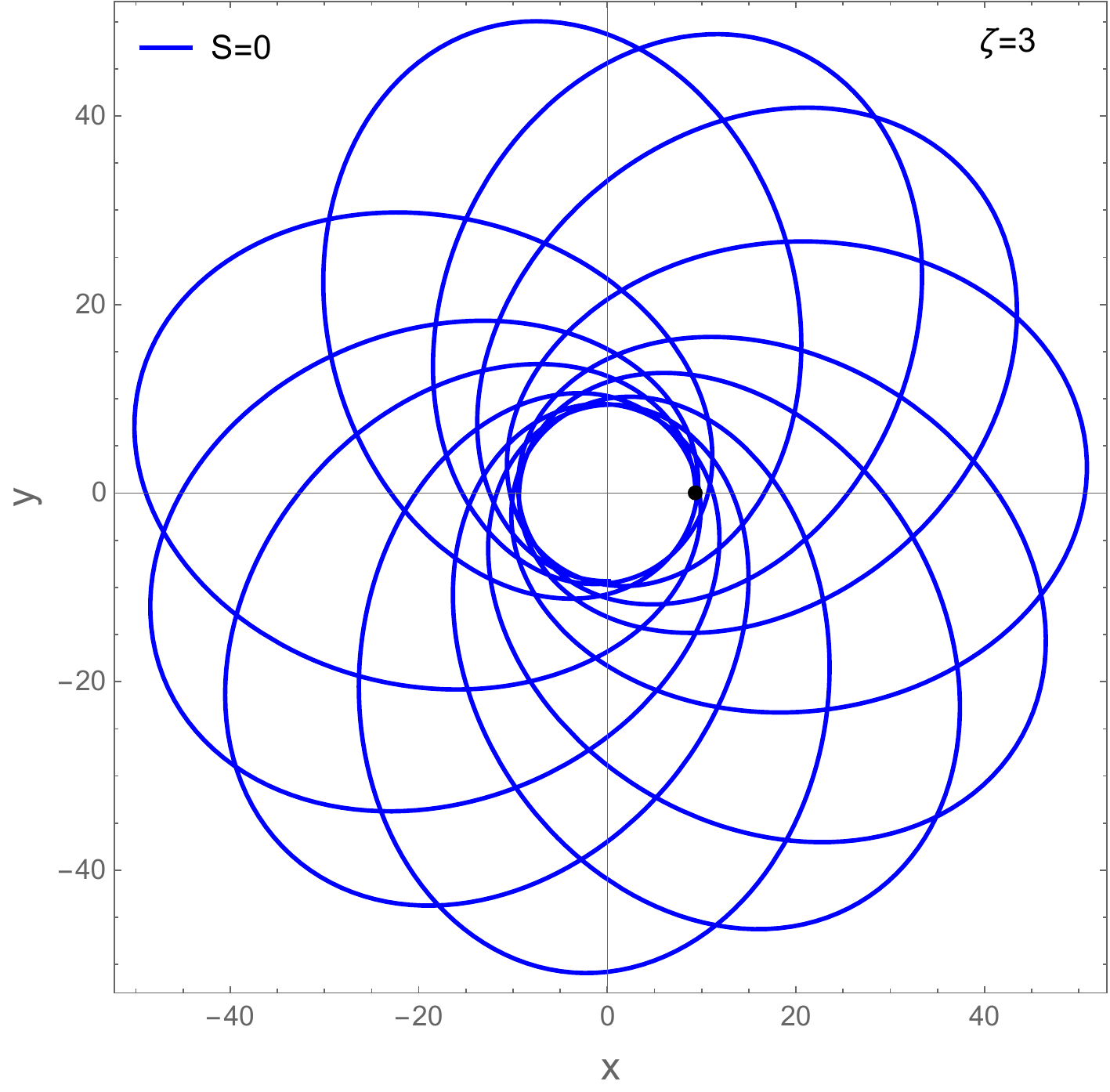}
		\caption{BH-III}
	\end{subfigure}
	\begin{subfigure}{0.33\textwidth}
		\includegraphics[height=5cm,keepaspectratio]{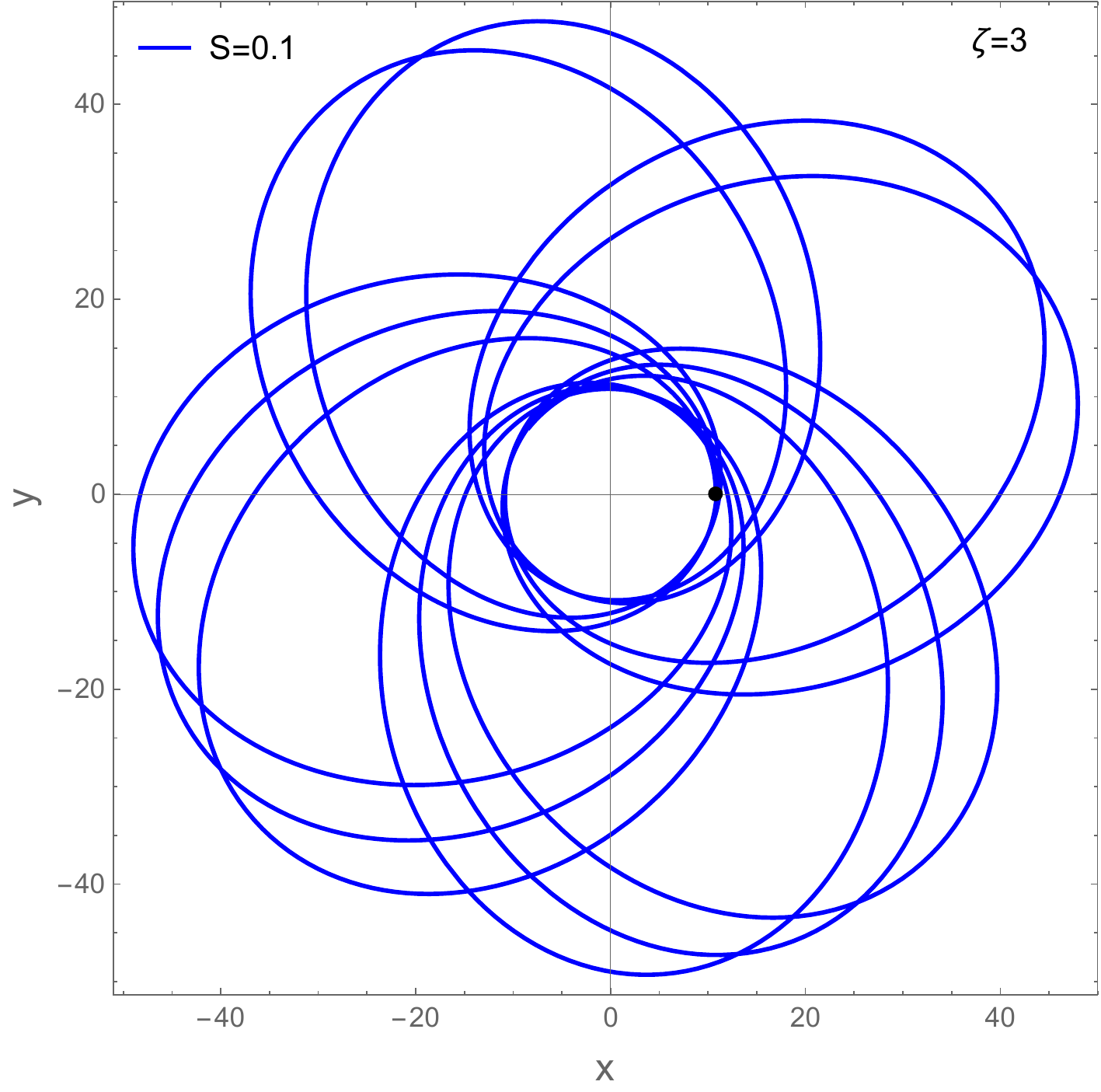}
		\caption{BH-III}
	\end{subfigure}

	\begin{subfigure}{0.33\textwidth}
		\includegraphics[height=5cm, keepaspectratio]{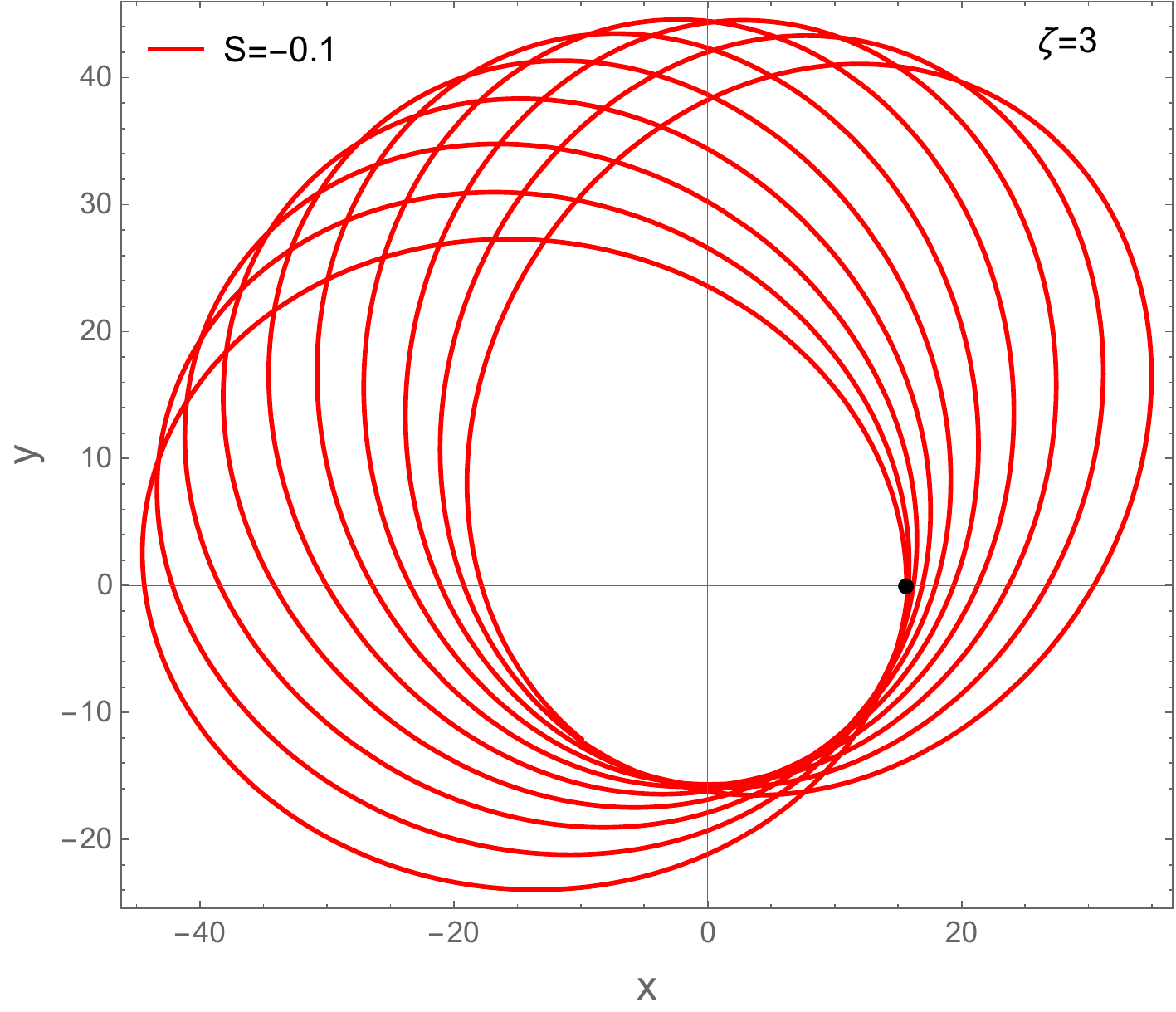}
		\caption{BH-I}
	\end{subfigure}
	\begin{subfigure}{0.33\textwidth}
		\includegraphics[height=5cm]{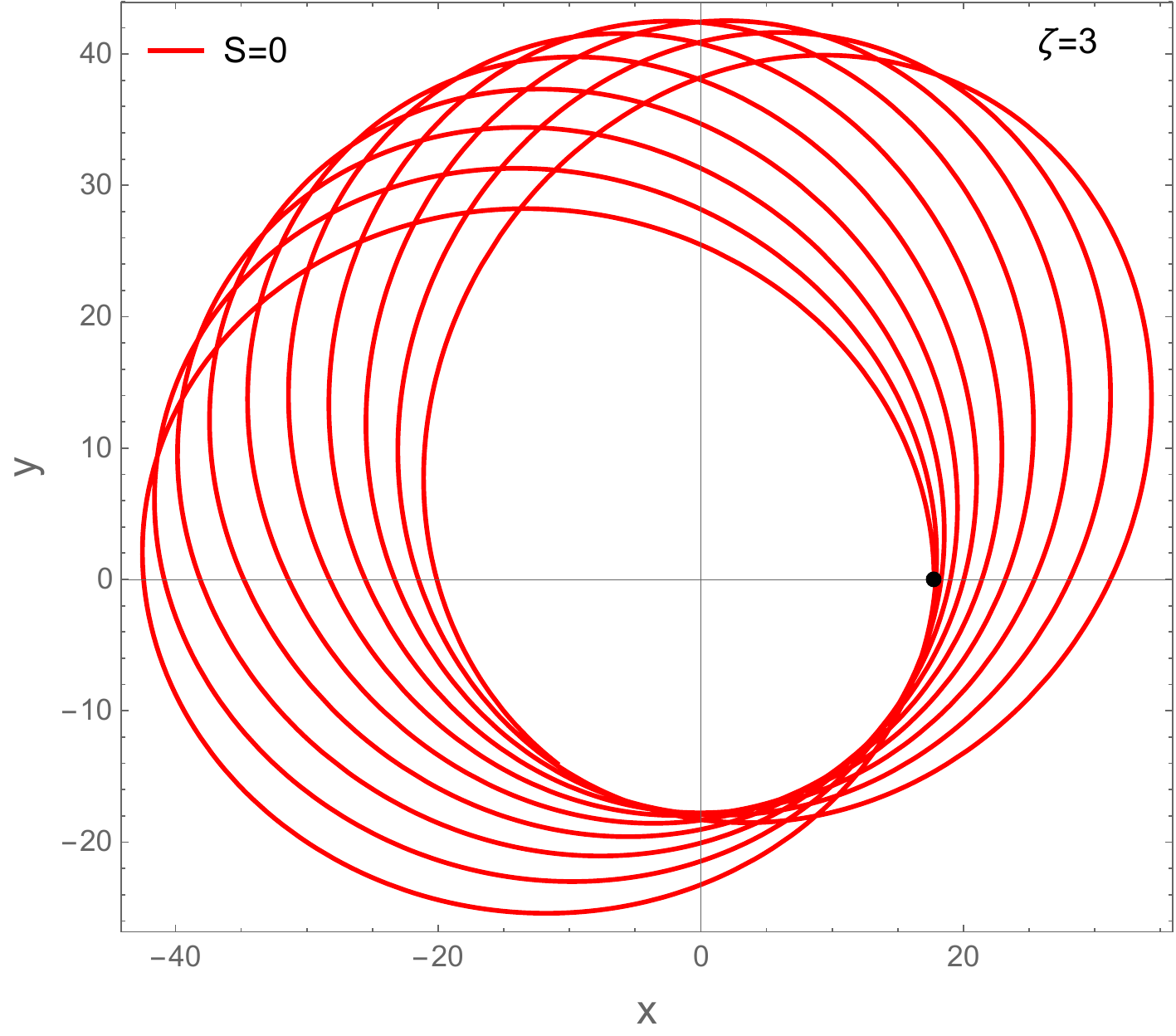}
		\caption{BH-I}
	\end{subfigure}
	\begin{subfigure}{0.33\textwidth}
		\includegraphics[height=5cm,keepaspectratio]{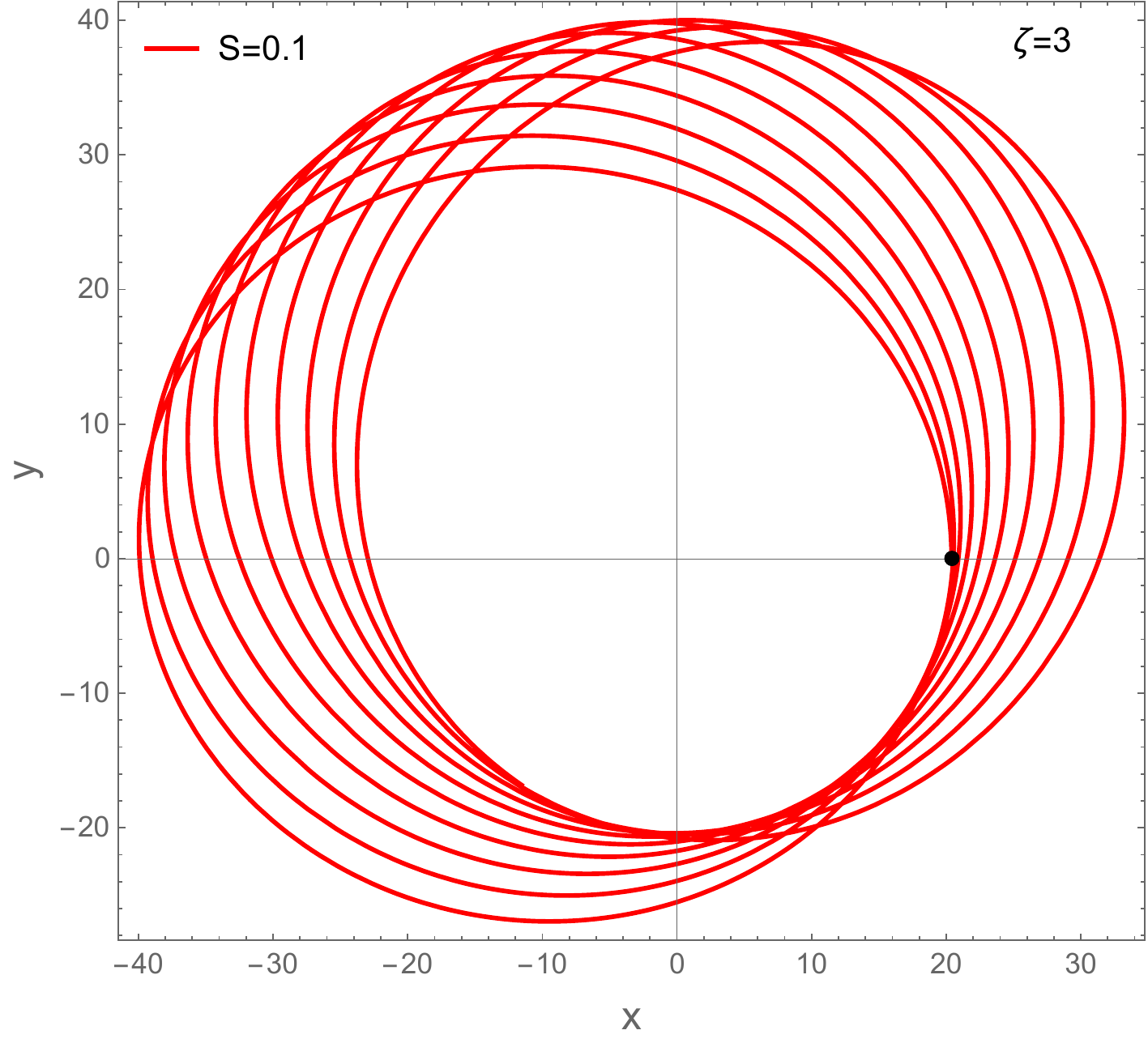}
		\caption{BH-I}
	\end{subfigure}
	\begin{subfigure}{0.33\textwidth}
		\includegraphics[height=5cm, keepaspectratio]{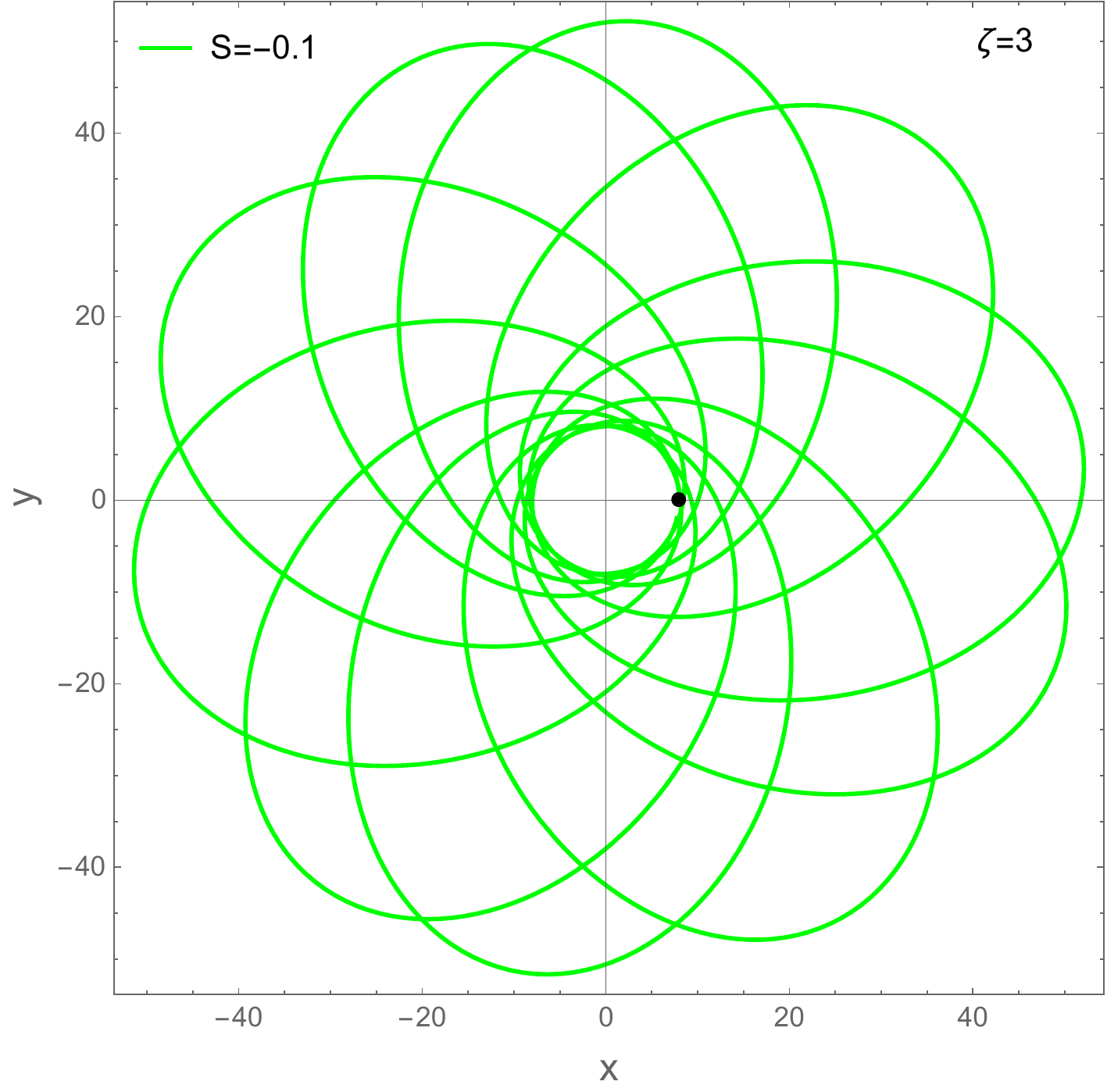}
		\caption{BH-II}
	\end{subfigure}
	\begin{subfigure}{0.33\textwidth}
		\includegraphics[height=5cm]{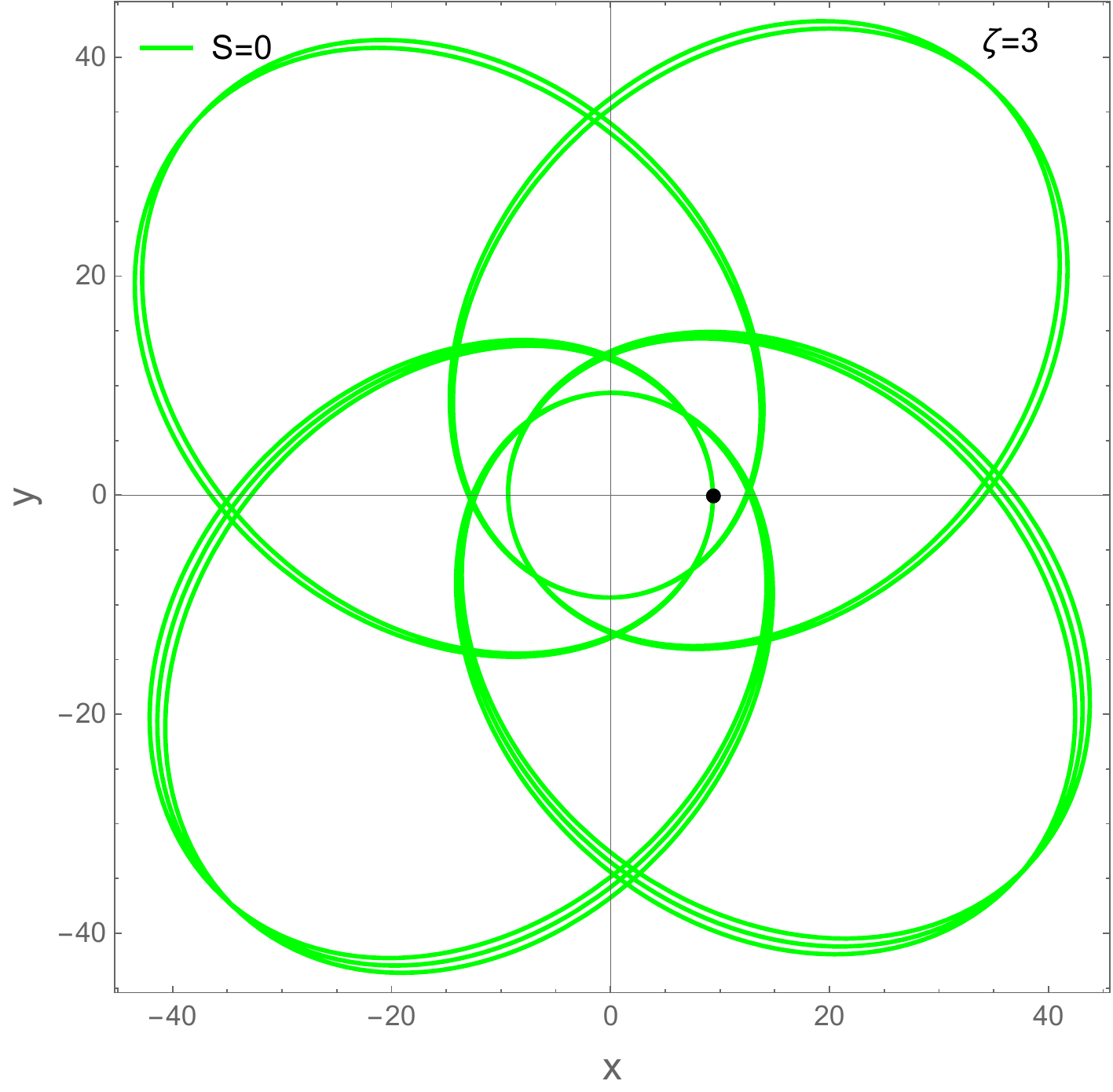}
		\caption{BH-II}
	\end{subfigure}
	\begin{subfigure}{0.33\textwidth}
		\includegraphics[height=5cm,keepaspectratio]{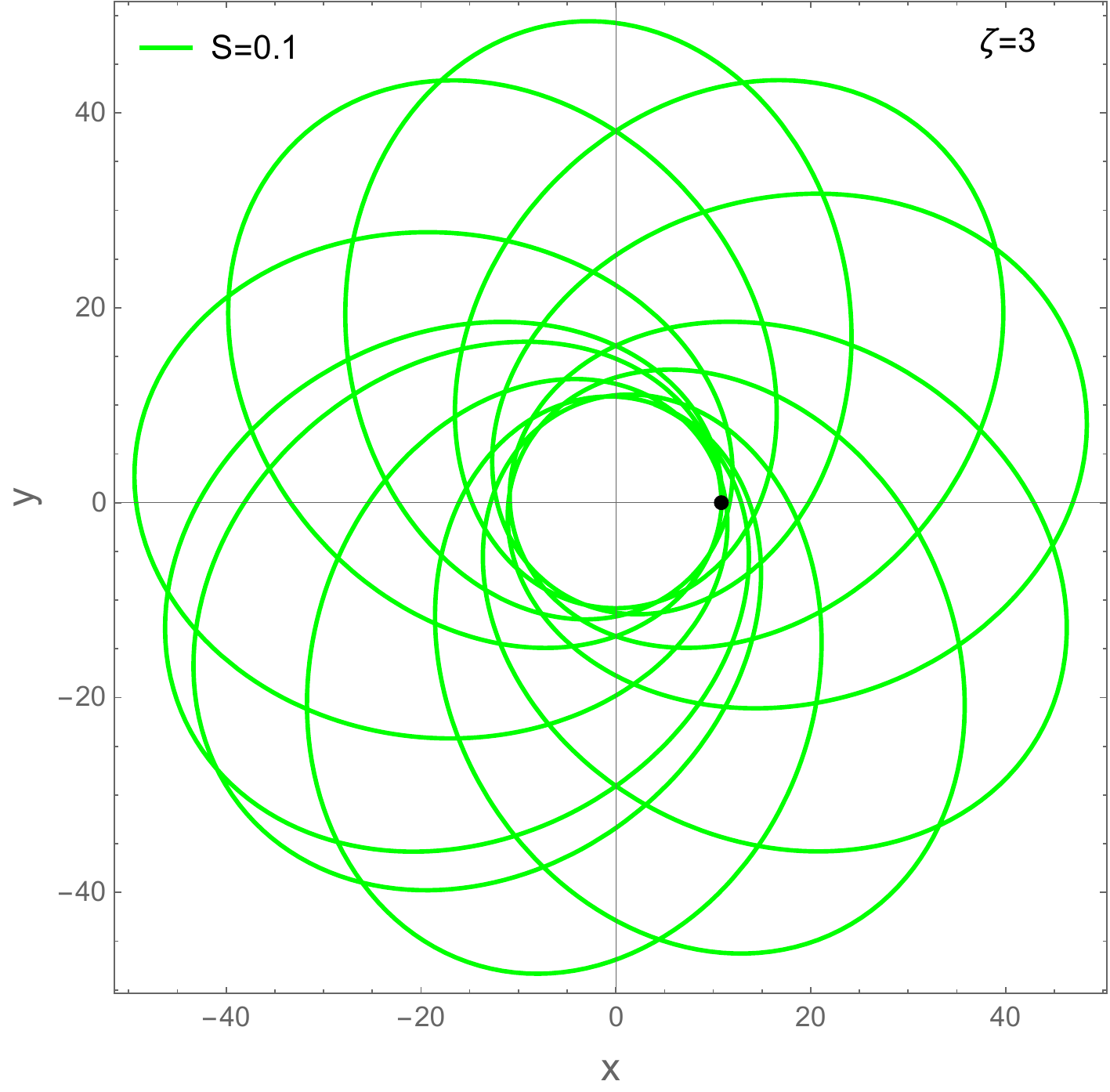}
		\caption{BH-II}
	\end{subfigure}
	\caption{Trajectories of spinning particles with $E=0.984$ and $L=4.5$ around three covariant quantum-corrected BHs ($\zeta=3$) and the Schwarzschild BH for different values of $S$.}
	\label{fig:guiji1}
\end{figure*}
To distinguish the recently proposed covariant quantum-corrected BH models, we present here the trajectories of spinning particles around the Schwarzschild BH, BH-I, BH-II, and BH-III, respectively. To satisfy the energy condition for bound orbits in all three BH models, we choose $L = 4.5$ and particle energy $E = 0.984$. The trajectories of spinning particles around each BH model are shown in Fig.~\ref{fig:guiji1}, where the black, blue, red, and green curves represent the Schwarzschild BH, BH-III, BH-I, and BH-II, respectively. The black dot denotes the starting point. Consistent with previous discussions, BH-III remains difficult to distinguish from the Schwarzschild BH under these initial conditions. However, the trajectories of spinning particles around BH-I and BH-II exhibit significant differences compared to BH-III. For BH-I, the periastron distance increases with increasing $S$, while the apastron distance decreases accordingly. It is noteworthy that in previous studies, the effective potential of BH-II does not vary with $\zeta$ when $S=0$~\cite{Umarov:2025wzm}, which is the same as that of the Schwarzschild case. Here, the same conclusion can be drawn; however, due to differences in the variation of the radial distance $r$ with $\phi$, the trajectories show noticeable deviations from the Schwarzschild case. In summary, we can distinguish BH-I, BH-II, and BH-III by examining the trajectories of spinning particles under specific initial conditions.

\section{Summary}\label{section5}

In this paper, we investigate the motion of test particles with intrinsic spin around a covariant quantum-corrected BH without a Cauchy horizon~\cite{Zhang:2024ney}. We first provide a brief review of this quantum-corrected BH (denoted as BH-III) and constrain the values of quantum parameter $\zeta$ used in this study based on the theoretical upper limit of the $\zeta$~\cite{Zhang:2024ney}.

Due to the spin-curvature interaction, the particle deviates from geodesic motion. Under the pole-dipole approximation, we employ the MPD equations to describe the motion of spinning particles in this quantum-corrected spacetime. Using certain spin supplementary conditions and conserved quantities generated by spacetime symmetries, we solve the MPD equations to obtain the 4-momentum and 4-velocity of the spinning particle. Subsequently, we derive the effective potential $V_{\rm eff}$ for the radial motion of the spinning particle from the components of the 4-momentum and analyze the influence of spin $S$ and quantum parameter $\zeta$ on $V_{\rm eff}$. We find that an increase in $\zeta$ leads to a decrease in $V_{\rm eff}$, and this effect of $\zeta$ is weaker compared to the influence of $S$ on $V_{\rm eff}$.

Furthermore, based on the effective potential $V_{\rm eff}$, we investigate the circular orbits of spinning particles and present in Fig.~\ref{fig_radius} the dependence of the circular orbit radius on the quantum parameter $\zeta$ and spin $S$. It can be observed that the influence of $S$ on the circular orbit radius is significantly greater than that of $\zeta$. We further analyze the behavior of ISCO under the influence of $S$ and $\zeta$, with the results shown in Fig.~\ref{fig:ISCO}. The study indicates that the presence of $\zeta$ always increases the ISCO-related physical quantities, but the increase is relatively small; whereas the influence of $S$ on these quantities is more pronounced. Due to the possible non-parallelism between the 4-velocity and 4-momentum of spinning particles, which may lead to the 4-velocity becoming spacelike, we discuss the timelike condition to avoid such unphysical situations. Taking the ISCO as an example, we display in Fig.~\ref{fig_time} the constraints on $S$ and $\zeta$ for particles at the ISCO satisfying the timelike condition.

Finally, we derive the equations of motion for spinning particles around this quantum-corrected black hole (BH-III). Given the initial conditions for bound orbits with $E=0.976$ and $L=4.5$, we obtain trajectories for different values of spin $S$ and quantum parameter $\zeta$ based on the equations of motion, as shown in Fig.~\ref{fig:guiji}. The results show that trajectories of spinning particles with different $S$ values in BH-III exhibit differences, while the influence of $\zeta$ on the trajectories is not significant. The spin dominates the motion of the particle, and when the spin is small, we cannot distinguish BH-III from the Schwarzschild BH through particle trajectories. In addition, we compare the trajectories of spinning particles in BH-III with those in two other covariant quantum-corrected BHs~\cite{Zhang:2024khj}, denoted as BH-I and BH-II (the dynamics of spinning particles in BH-I and BH-II have been previously studied~\cite{Du:2024ujg,Umarov:2025wzm}). In Fig.~\ref{fig:guiji1}, we observe that for spinning particles with $E=0.984$ and $L=4.5$, their bound orbital trajectories around these three quantum-corrected BHs are distinct, enabling us to differentiate between BH-I, BH-II, and BH-III through these trajectories.

In conclusion, this study extends previous research on the motion of spinning particles~\cite{Du:2024ujg,Umarov:2025wzm} in covariant quantum-corrected BH models to the case of a covariant quantum-corrected BH model without a Cauchy horizon, BH-III, and provides a brief comparison with two other quantum-corrected models, BH-I and BH-II. Our results reveal that there may exist differences in the trajectories of spinning particles on bound orbits around these BHs. This may provide a potential tool for future studies on the motion of spinning particles around quantum-corrected BHs and for distinguishing between different models.
\begin{acknowledgments}
This work is supported in part by NSFC Grant No. 12165005.
\end{acknowledgments}





\begin{thebibliography}{65}

\bibitem{Penrose:1964wq}
R.~Penrose, {Gravitational Collapse and Space-time Singularities}. Phys. Rev. Lett. \textbf{14}, 57 (1965). {\url{https://doi.org/10.1103/PhysRevLett.14.57}}

\bibitem{Hawking:1970zqf}
S.W.~Hawking and R.~Penrose, {The Singularities of gravitational collapse and cosmology}. Proc. Roy. Soc. Lond. A \textbf{314}, 529 (1970). {\url{https://doi.org/10.1098/rspa.1970.0021}}

\bibitem{Surya:2019ndm}
S.~Surya, {The causal set approach to quantum gravity}. Living Rev. Rel. \textbf{22}, 5 (2019). {\url{https://doi.org/10.1007/s41114-019-0023-1}}. {\href{https://arxiv.org/abs/1903.11544}{{arXiv:1903.11544}}}

\bibitem{Rovelli:2011eq}
C.~Rovelli, {Zakopane lectures on loop gravity}. PoS \textbf{QGQGS2011}, 003 (2011). {\url{https://doi.org/10.22323/1.140.0003}}. {\href{https://arxiv.org/abs/1102.3660}{{arXiv:1102.3660}}}

\bibitem{Rovelli:1997yv}
C.~Rovelli, {Loop quantum gravity}. Living Rev. Rel. \textbf{1}, 1 (1998). {\url{https://doi.org/10.12942/lrr-1998-1}}. {\href{https://arxiv.org/abs/gr-qc/9710008}{{arXiv:gr-qc/9710008}}}

\bibitem{Ashtekar:2005qt}
A.~Ashtekar and M.~Bojowald, {Quantum geometry and the Schwarzschild singularity}. Class. Quant. Grav. \textbf{23}, 391 (2006). {\url{https://doi.org/10.1088/0264-9381/23/2/008}}. {\href{https://arxiv.org/abs/gr-qc/0509075}{{arXiv:gr-qc/0509075}}}

\bibitem{Ashtekar:2004eh}
A.~Ashtekar and J.~Lewandowski, {Background independent quantum gravity: A Status report}. Class. Quant. Grav. \textbf{21}, R53 (2004). {\url{https://doi.org/10.1088/0264-9381/21/15/R01}}. {\href{https://arxiv.org/abs/gr-qc/0404018}{{arXiv:gr-qc/0404018}}}

\bibitem{Ashtekar:2013hs}
A.~Ashtekar, {Introduction to loop quantum gravity and cosmology}. Lect. Notes Phys. \textbf{863}, 31 (2013). {\url{https://doi.org/10.1007/978-3-642-33036-0_2}}. {\href{https://arxiv.org/abs/1201.4598}{{arXiv:1201.4598}}}

\bibitem{Han:2005km}
M.~Han, W.~Huang, and Y.~Ma, {Fundamental structure of loop quantum gravity}. Int. J. Mod. Phys. D \textbf{16}, 1397 (2007). {\url{https://doi.org/10.1142/S0218271807010894}}. {\href{https://arxiv.org/abs/gr-qc/0509064}{{arXiv:gr-qc/0509064}}}

\bibitem{Modesto:2008im}
L.~Modesto, {Semiclassical loop quantum black hole}. Int. J. Theor. Phys. \textbf{49}, 1649 (2010). {\url{https://doi.org/10.1007/s10773-010-0346-x}}. {\href{https://arxiv.org/abs/0811.2196}{{arXiv:0811.2196}}}

\bibitem{Perez:2017cmj}
A.~Perez, {Black Holes in Loop Quantum Gravity}. Rept. Prog. Phys. \textbf{80}, 126901 (2017). {\url{https://doi.org/10.1088/1361-6633/aa7e14}}. {\href{https://arxiv.org/abs/1703.09149}{{arXiv:1703.09149}}}

\bibitem{Ashtekar:2018lag}
A.~Ashtekar, J.~Olmedo, and P.~Singh, {Quantum Transfiguration of Kruskal Black Holes}. Phys. Rev. Lett. \textbf{121}, 241301 (2018). {\url{https://doi.org/10.1103/PhysRevLett.121.241301}}. {\href{https://arxiv.org/abs/1806.00648}{{arXiv:1806.00648}}}

\bibitem{Bodendorfer:2019cyv}
N.~Bodendorfer, F.M.~Mele, and J.~M\"unch, {Effective Quantum Extended Spacetime of Polymer Schwarzschild Black Hole}. Class. Quant. Grav. \textbf{36}, 195015 (2019). {\url{https://doi.org/10.1088/1361-6382/ab3f16}}. {\href{https://arxiv.org/abs/1902.04542}{{arXiv:1902.04542}}}

\bibitem{Kelly:2020uwj}
J.G.~Kelly, R.~Santacruz, and E.~Wilson-Ewing, {Effective loop quantum gravity framework for vacuum spherically symmetric spacetimes}. Phys. Rev. D \textbf{102}, 106024 (2020). {\url{https://doi.org/10.1103/PhysRevD.102.106024}}. {\href{https://arxiv.org/abs/2006.09302}{{arXiv:2006.09302}}}

\bibitem{Gan:2020dkb}
W.-C.~Gan, N.O.~Santos, F.-W.~Shu, and A.~Wang, {Properties of the spherically symmetric polymer black holes}. Phys. Rev. D \textbf{102}, 124030 (2020). {\url{https://doi.org/10.1103/PhysRevD.102.124030}}. {\href{https://arxiv.org/abs/2008.09664}{{arXiv:2008.09664}}}

\bibitem{Sartini:2020ycs}
F.~Sartini and M.~Geiller, {Quantum dynamics of the black hole interior in loop quantum cosmology}. Phys. Rev. D \textbf{103}, 066014 (2021). {\url{https://doi.org/10.1103/PhysRevD.103.066014}}. {\href{https://arxiv.org/abs/2010.07056}{{arXiv:2010.07056}}}

\bibitem{Song:2020arr}
S.~Song, H.~Li, Y.~Ma, and C.~Zhang, {Entropy of black holes with arbitrary shapes in loop quantum gravity}. Sci. China Phys. Mech. Astron. \textbf{64}, 120411 (2021). {\url{https://doi.org/10.1007/s11433-021-1770-3}}. {\href{https://arxiv.org/abs/2002.08869}{{arXiv:2002.08869}}}

\bibitem{Zhang:2020qxw}
C.~Zhang, Y.~Ma, S.~Song, and X.~Zhang, {Loop quantum Schwarzschild interior and black hole remnant}. Phys. Rev. D \textbf{102}, 041502 (2020). {\url{https://doi.org/10.1103/PhysRevD.102.041502}}. {\href{https://arxiv.org/abs/2006.08313}{{arXiv:2006.08313}}}

\bibitem{Zhang:2021wex}
C.~Zhang, Y.~Ma, S.~Song, and X.~Zhang, {Loop quantum deparametrized Schwarzschild interior and discrete black hole mass}. Phys. Rev. D \textbf{105}, 024069 (2022). {\url{https://doi.org/10.1103/PhysRevD.105.024069}}. {\href{https://arxiv.org/abs/2107.10579}{{arXiv:2107.10579}}}

\bibitem{Lewandowski:2022zce}
J.~Lewandowski, Y.~Ma, J.~Yang, and C.~Zhang, {Quantum Oppenheimer-Snyder and Swiss Cheese Models}. Phys. Rev. Lett. \textbf{130}, 101501 (2023). {\url{https://doi.org/10.1103/PhysRevLett.130.101501}}. {\href{https://arxiv.org/abs/2210.02253}{{arXiv:2210.02253}}}

\bibitem{Bojowald:2015zha}
M.~Bojowald, S.~Brahma, and J.D.~Reyes, {Covariance in models of loop quantum gravity: Spherical symmetry}. Phys. Rev. D \textbf{92}, 045043 (2015). {\url{https://doi.org/10.1103/PhysRevD.92.045043}}. {\href{https://arxiv.org/abs/1507.00329}{{arXiv:1507.00329}}}

\bibitem{Bojowald:2015sta}
M.~Bojowald and S.~Brahma, {Covariance in models of loop quantum gravity: Gowdy systems}. Phys. Rev. D \textbf{92}, 065002 (2015). {\url{https://doi.org/10.1103/PhysRevD.92.065002}}. {\href{https://arxiv.org/abs/1507.00679}{{arXiv:1507.00679}}}

\bibitem{BenAchour:2017jof}
J.~Ben~Achour and S.~Brahma, {Covariance in self dual inhomogeneous models of effective quantum geometry: Spherical symmetry and Gowdy systems}. Phys. Rev. D \textbf{97}, 126003 (2018). {\url{https://doi.org/10.1103/PhysRevD.97.126003}}. {\href{https://arxiv.org/abs/1712.03677}{{arXiv:1712.03677}}}

\bibitem{Bojowald:2020unm}
M.~Bojowald, {No-go result for covariance in models of loop quantum gravity}. Phys. Rev. D \textbf{102}, 046006 (2020). {\url{https://doi.org/10.1103/PhysRevD.102.046006}}. {\href{https://arxiv.org/abs/2007.16066}{{arXiv:2007.16066}}}

\bibitem{Gambini:2022dec}
R.~Gambini, J.~Olmedo, and J.~Pullin, {Towards a quantum notion of covariance in spherically symmetric loop quantum gravity}. Phys. Rev. D \textbf{105}, 026017 (2022). {\url{https://doi.org/10.1103/PhysRevD.105.026017}}. {\href{https://arxiv.org/abs/2201.01616}{{arXiv:2201.01616}}}

\bibitem{Han:2022rsx}
M.~Han and H.~Liu, {Covariant \ensuremath{\bar\mu}-scheme effective dynamics, mimetic gravity, and nonsingular black holes: Applications to spherically symmetric quantum gravity}. Phys. Rev. D \textbf{109}, 084033 (2024). {\url{https://doi.org/10.1103/PhysRevD.109.084033}}. {\href{https://arxiv.org/abs/2212.04605}{{arXiv:2212.04605}}}

\bibitem{Li:2023axl}
S.~Li and J.-P.~Wu, {Gravitational waves with generalized holonomy corrections}. Eur. Phys. J. C \textbf{84}, 629 (2024). {\url{https://doi.org/10.1140/epjc/s10052-024-13010-2}}. {\href{https://arxiv.org/abs/2309.05535}{{arXiv:2309.05535}}}

\bibitem{Zhang:2024khj}
C.~Zhang, J.~Lewandowski, Y.~Ma, and J.~Yang, {Black holes and covariance in effective quantum gravity}. Phys. Rev. D \textbf{111}, L081504 (2025). {\url{https://doi.org/10.1103/PhysRevD.111.L081504}}. {\href{https://arxiv.org/abs/2407.10168}{{arXiv:2407.10168}}}

\bibitem{Zhang:2024ney}
C.~Zhang, J.~Lewandowski, Y.~Ma, and J.~Yang, {Black holes and covariance in effective quantum gravity: A solution without Cauchy horizons}. Phys. Rev. D \textbf{112}, 044054 (2025). {\url{https://doi.org/10.1103/d6ks-d576}}. {\href{https://arxiv.org/abs/2412.02487}{{arXiv:2412.02487}}}

\bibitem{Yang:2025ufs}
J.~Yang, C.~Zhang, and Y.~Ma, {Covariant effective spacetimes of spherically symmetric electro-vacuum with a cosmological constant}. {\href{https://arxiv.org/abs/2503.15157}{{arXiv:2503.15157}}}

\bibitem{Konoplya:2024lch}
R.A.~Konoplya and O.S.~Stashko, {Probing the effective quantum gravity via quasinormal modes and shadows of black holes}. Phys. Rev. D \textbf{111}, 104055 (2025). {\url{https://doi.org/10.1103/PhysRevD.111.104055}}. {\href{https://arxiv.org/abs/2408.02578}{{arXiv:2408.02578}}}

\bibitem{Liu:2024soc}
W.~Liu, D.~Wu, and J.~Wang, {Light rings and shadows of static black holes in effective quantum gravity}. Phys. Lett. B \textbf{858}, 139052 (2024). {\url{https://doi.org/10.1016/j.physletb.2024.139052}}. {\href{https://arxiv.org/abs/2408.05569}{{arXiv:2408.05569}}}

\bibitem{Liu:2024wal}
H.~Liu, M.-Y.~Lai, X.-Y.~Pan, H.~Huang, and D.-C.~Zou, {Gravitational lensing effect of black holes in effective quantum gravity}. Phys. Rev. D \textbf{110}, 104039 (2024). {\url{https://doi.org/10.1103/PhysRevD.110.104039}}. {\href{https://arxiv.org/abs/2408.11603}{{arXiv:2408.11603}}}

\bibitem{Zhu:2024wic}
L.-G.~Zhu, G.~Fu, S.~Li, D.~Zhang, and J.-P.~Wu, {Quasinormal modes of a charged loop quantum black hole}. Phys. Rev. D \textbf{111}, 104008 (2025). {\url{https://doi.org/10.1103/PhysRevD.111.104008}}. {\href{https://arxiv.org/abs/2410.00543}{{arXiv:2410.00543}}}

\bibitem{Wang:2024iwt}
Y.~Wang, A.~Vachher, Q.~Wu, T.~Zhu, and S.G.~Ghosh, {Strong gravitational lensing by static black holes in effective quantum gravity}. Eur. Phys. J. C \textbf{85}, 302 (2025). {\url{https://doi.org/10.1140/epjc/s10052-025-13970-z}}. {\href{https://arxiv.org/abs/2410.12382}{{arXiv:2410.12382}}}

\bibitem{Ban:2024qsa}
Z.~Ban, J.~Chen, and J.~Yang, {Shadows of rotating black holes in effective quantum gravity}. Eur. Phys. J. C \textbf{85}, 878 (2025). {\url{https://doi.org/10.1140/epjc/s10052-025-14614-y}}. {\href{https://arxiv.org/abs/2411.09374}{{arXiv:2411.09374}}}

\bibitem{Lin:2024beb}
J.~Lin, X.~Zhang, and M.~Bravo-Gaete, {Mass inflation and strong cosmic censorship conjecture in the covariant quantum black hole}. Phys. Rev. D \textbf{111}, 106025 (2025). {\url{https://doi.org/10.1103/n7jv-crs9}}. {\href{https://arxiv.org/abs/2412.01448}{{arXiv:2412.01448}}}

\bibitem{Shu:2024tut}
Y.-H.~Shu and J.-H.~Huang, {Circular orbits and thin accretion disk around a quantum corrected black hole}. Phys. Lett. B \textbf{864}, 139411 (2025). {\url{https://doi.org/10.1016/j.physletb.2025.139411}}. {\href{https://arxiv.org/abs/2412.05670}{{arXiv:2412.05670}}}

\bibitem{Liu:2024iec}
W.~Liu, D.~Wu, and J.~Wang, {Light rings and shadows of static black holes in effective quantum gravity II: A new solution without Cauchy horizons}. Phys. Lett. B \textbf{868}, 139742 (2025). {\url{https://doi.org/10.1016/j.physletb.2025.139742}}. {\href{https://arxiv.org/abs/2412.18083}{{arXiv:2412.18083}}}

\bibitem{Bojowald:2024ium}
M.~Bojowald, E.I.~Duque, and D.~Hartmann, {Covariant Lema\^{\i}tre-Tolman-Bondi collapse in models of loop quantum gravity}. Phys. Rev. D \textbf{111}, 064002 (2025). {\url{https://doi.org/10.1103/PhysRevD.111.064002}}. {\href{https://arxiv.org/abs/2412.18054}{{arXiv:2412.18054}}}

\bibitem{Konoplya:2025hgp}
R.A.~Konoplya and O.S.~Stashko, {Transition from regular black holes to wormholes in covariant effective quantum gravity: Scattering, quasinormal modes, and Hawking radiation}. Phys. Rev. D \textbf{111}, 084031 (2025). {\url{https://doi.org/10.1103/PhysRevD.111.084031}}. {\href{https://arxiv.org/abs/2502.05689}{{arXiv:2502.05689}}}

\bibitem{Chen:2025ifv}
J.~Chen and J.~Yang, {Shadows and optical appearance of quantum-corrected black holes illuminated by static thin accretions}. Eur. Phys. J. C \textbf{85}, 512 (2025). {\url{https://doi.org/10.1140/epjc/s10052-025-14230-w}}. {\href{https://arxiv.org/abs/2503.06215}{{arXiv:2503.06215}}}

\bibitem{Lutfuoglu:2025hwh}
B.C.~L{\"u}tf{\"u}o{\u{g}}lu, {Long-lived quasinormal modes around regular black holes and wormholes in Covariant Effective Quantum Gravity}. JCAP \textbf{06}, 057 (2025). {\url{https://doi.org/10.1088/1475-7516/2025/06/057}}. {\href{https://arxiv.org/abs/2504.09323}{{arXiv:2504.09323}}}

\bibitem{Chen:2025aqh}
J.~Chen and J.~Yang, {Periodic orbits and gravitational waveforms in quantum-corrected black hole spacetimes}. Eur. Phys. J. C \textbf{85}, 726 (2025). {\url{https://doi.org/10.1140/epjc/s10052-025-14457-7}}. {\href{https://arxiv.org/abs/2505.02660}{{arXiv:2505.02660}}}

\bibitem{Al-Badawi:2025rcq}
A.~Al-Badawi, F.~Ahmed, and {\.I}.~Sakall{\i}, {Effective quantum gravity black hole with cloud of strings surrounded by quintessence field}. Nucl. Phys. B \textbf{1017}, 116961 (2025). {\url{https://doi.org/10.1016/j.nuclphysb.2025.116961}}

\bibitem{Zhang:2025ccx}
C.~Zhang and Z.~Cao, {Covariant dynamics from static spherically symmetric geometries}. {\href{https://arxiv.org/abs/2506.09540}{{arXiv:2506.09540}}}

\bibitem{Sekhmani:2025bsi}
Y.~Sekhmani, H.~Ali, S.G.~Ghosh, and K.~Boshkayev, {Rotating charged nonsingular black holes in loop quantum gravity and their observational imprints from EHT}. JHEAp \textbf{49}, 100425 (2026). {\url{https://doi.org/10.1016/j.jheap.2025.100425}}

\bibitem{LIGO:2017dbh}
B.P.~Abbott {\em et~al.} (LIGO Scientific Collaboration and Virgo Collaboration), {Observation of Gravitational Waves from a Binary Black Hole Merger}, in \emph{{Centennial of General Relativity}: {A Celebration}}, edited by C.A.Z.~Vasconcellos (World Scientific, Singapore, 2017). {\url{https://doi.org/10.1142/9789814699662_0011}}

\bibitem{EventHorizonTelescope:2019dse}
K.~Akiyama {\em et~al.} (Event Horizon Telescope Collaboration), {First M87 Event Horizon Telescope Results. I. The Shadow of the Supermassive Black Hole}. Astrophys. J. Lett. \textbf{875}, L1 (2019). {\url{https://doi.org/10.3847/2041-8213/ab0ec7}}. {\href{https://arxiv.org/abs/1906.11238}{{arXiv:1906.11238}}}

\bibitem{Hughes:2000ssa}
S.A.~Hughes, {Gravitational waves from extreme mass ratio inspirals: Challenges in mapping the space-time of massive, compact objects}. Class. Quant. Grav. \textbf{18}, 4067 (2001). {\url{https://doi.org/10.1088/0264-9381/18/19/314}}. {\href{https://arxiv.org/abs/gr-qc/0008058}{{arXiv:gr-qc/0008058}}}

\bibitem{Jefremov:2015gza}
P.I.~Jefremov, O.Y.~Tsupko, and G.S.~Bisnovatyi-Kogan, {Innermost stable circular orbits of spinning test particles in Schwarzschild and Kerr space-times}. Phys. Rev. D \textbf{91}, 124030 (2015). {\url{https://doi.org/10.1103/PhysRevD.91.124030}}. {\href{https://arxiv.org/abs/1503.07060}{{arXiv:1503.07060}}}

\bibitem{Mathisson:1937zz}
M.~Mathisson, {Neue mechanik materieller systemes}. Acta Phys. Polon. \textbf{6}, 163 (1937)

\bibitem{Papapetrou:1951pa}
A.~Papapetrou, {Spinning test particles in general relativity. 1.}
Proc. Roy. Soc. Lond. A \textbf{209}, 248 (1951). {\url{https://doi.org/10.1098/rspa.1951.0200}}

\bibitem{Dixon:1970zza}
W.G.~Dixon, {Dynamics of extended bodies in general relativity. I. Momentum and angular momentum}. Proc. Roy. Soc. Lond. A \textbf{314}, 499 (1970). {\url{https://doi.org/10.1098/rspa.1970.0020}}

\bibitem{Dixon:1970zz}
W.G.~Dixon, {Dynamics of extended bodies in general relativity. II. Moments of the charge-current vector}. Proc. Roy. Soc. Lond. A \textbf{319}, 509 (1970). {\url{https://doi.org/10.1098/rspa.1970.0191}}

\bibitem{Dixon:1974xoz}
W.G.~Dixon, {Dynamics of extended bodies in general relativity III. Equations of motion}. Phil. Trans. Roy. Soc. Lond. A \textbf{277}, 59 (1974). {\url{https://doi.org/10.1098/rsta.1974.0046}}

\bibitem{Zhang:2017nhl}
Y.-P.~Zhang, S.-W.~Wei, W.-D.~Guo, T.-T.~Sui, and Y.-X.~Liu, {Innermost stable circular orbit of spinning particle in charged spinning black hole background}. Phys. Rev. D \textbf{97}, 084056 (2018). {\url{https://doi.org/10.1103/PhysRevD.97.084056}}. {\href{https://arxiv.org/abs/1711.09361}{{arXiv:1711.09361}}}

\bibitem{Toshmatov:2019bda}
B.~Toshmatov and D.~Malafarina, {Spinning test particles in the $\gamma$ spacetime}. Phys. Rev. D \textbf{100}, 104052 (2019). {\url{https://doi.org/10.1103/PhysRevD.100.104052}}. {\href{https://arxiv.org/abs/1910.11565}{{arXiv:1910.11565}}}

\bibitem{Zhang:2022qzw}
Y.-P.~Zhang, Y.-B.~Zeng, Y.-Q.~Wang, S.-W.~Wei, and Y.-X.~Liu, {Equatorial orbits of spinning test particles in rotating boson stars}. Eur. Phys. J. C \textbf{82}, 809 (2022). {\url{https://doi.org/10.1140/epjc/s10052-022-10743-w}}. {\href{https://arxiv.org/abs/2201.01498}{{arXiv:2201.01498}}}

\bibitem{Rakhimova:2024hzt}
G.~Rakhimova, F.~Atamurotov, N.~Juraeva, A.~Abdujabbarov, and G.~Mustafa, {Spinning particle motion around charged decoupled hairy black hole}. Phys. Dark Univ. \textbf{47}, 101721 (2025). {\url{https://doi.org/10.1016/j.dark.2024.101721}}

\bibitem{Tan:2024hzw}
Q.~Tan, W.~Deng, S.~Long, and J.~Jing, {Motion of spinning particles around black hole in a dark matter halo}. JCAP \textbf{05}, 044 (2025). {\url{https://doi.org/10.1088/1475-7516/2025/05/044}}. {\href{https://arxiv.org/abs/2409.17760}{{arXiv:2409.17760}}}

\bibitem{Skoupy:2024uan}
V.~Skoup{\'y} and V.~Witzany, {Analytic Solution for the Motion of Spinning Particles in Kerr Spacetime}. Phys. Rev. Lett. \textbf{134}, 171401 (2025). {\url{https://doi.org/10.1103/PhysRevLett.134.171401}}. {\href{https://arxiv.org/abs/2411.16855}{{arXiv:2411.16855}}}

\bibitem{Alimova:2025izs}
A.~Alimova, F.~Atamurotov, A.~Abdujabbarov, G.~Mustafa, and P.~Channuie, {Impact of quantum-corrected parameter on spinning particle motion around a black hole}. Eur. Phys. J. C \textbf{85}, 646 (2025). {\url{https://doi.org/10.1140/epjc/s10052-025-14385-6}}

\bibitem{Du:2024ujg}
Y.~Du, Y.~Liu, and X.~Zhang, {Spinning particle dynamics and the innermost stable circular orbit in covariant loop quantum gravity}. JCAP \textbf{05}, 045 (2025). {\url{https://doi.org/10.1088/1475-7516/2025/05/045}}. {\href{https://arxiv.org/abs/2411.13316}{{arXiv:2411.13316}}}

\bibitem{Umarov:2025wzm}
D.~Umarov, F.~Atamurotov, S.G.~Ghosh, A.~Abdujabbarov, and G.~Mustafa, {Dynamics of spinning particles around static black holes in effective quantum gravity}. Eur. Phys. J. C \textbf{85}, 800 (2025). {\url{https://doi.org/10.1140/epjc/s10052-025-14541-y}}

\end{thebibliography}
\end{document}